# Gamma Factory high-intensity muon and positron source: Exploratory studies

Armen Apyan,[1,*] Mieczyslaw Witold Krasny,[2,3] and Wiesław Płaczek[4]

[1]*A. Alikhanyan National Laboratory (AANL),
2 Alikhanian Brothers St., 0036 Yerevan, Armenia*
[2]*LPNHE, Sorbonne Université, Université de Paris, CNRS/IN2P3,
Tour 33, RdC, 4, pl. Jussieu, 75005 Paris, France*
[3]*CERN, BE-ABP, 1211 Geneva 23, Switzerland*
[4]*Institute of Applied Computer Science and Mark Kac Center for Complex Systems Research,
Jagiellonian University, ul. Łojasiewicza 11, 30-348 Kraków, Poland*

One of the fundamental challenges for future leptonic colliders and neutrino factories as well as for high-sensitivity studies of lepton universality is to design and construct new high-intensity sources of muons and positrons. The next-generation sources should increase the intensity of the presently operating ones by at least three orders of magnitude and include an important option of producing longitudinally polarized leptons. The main effort to achieve this goal has been focused, so far, on the proton-beam-driven muon sources and electron-beam-driven positron sources. In this paper, we present exploratory studies of an alternative scheme which is based on high-intensity megawatt-class photon beams. Such beams could be delivered in the future by the Gamma Factory (GF) project. One of the GF multiple goals is to increase the energy range (by more than one order of magnitude) and the intensity (by more than six orders of magnitude) of presently operating photon sources. Such a leap can be achieved by extending the present hadron-collider *modus operandi* of the LHC with a new *GF-operation-mode*, allowing the collisions of beams with laser pulses. The exploratory studies presented in this paper demonstrate that more than $10^{13}$ muons of both signs and more than $10^{16}$ electrons/positrons per second can be produced by a GF source.

## I. INTRODUCTION

All currently operating muon sources are driven by proton beams. The proton beams impinge on stationary targets producing pions and kaons. Their decays produce muons.

The high-intensity sources at PSI [1], J-PARC [2], FNAL [3] deliver up to $4 \times 10^8$ muons per second[1]. The goal of their planned upgrades is to reach intensities of $4 \times 10^{10}$ muons/s (PSI) [4] and $2 \times 10^{12}$ muons/s (FNAL) [5]. None of these proton-beam-driven sources satisfy the intensity requirements of the muon collider and neutrino factory projects [6, 7]. These projects require designing and constructing new proton accelerators, and extensive R&D on the target technology for megawatt proton beams to prove that the muon production rate in the range $10^{13}$–$10^{14}$ of both $\mu^+$ and $\mu^-$ per second can be achieved.

The source-intensity requirement for the energy-frontier muon collider becomes less important in a recently proposed, alternative, positron-beam-driven muon production method called the LEMMA scheme [8]. The concept of this scheme is based on the muon-beam-intensity versus muon-beam-emittance trade-off – a less intense source at proportionally smaller initial emittance of the muon beam. Muons in the LEMMA scheme are proposed to be produced in collisions of the high-energy positron beam with stationary-target plasma electrons [8]. The requisite positron source intensity needed for this scheme, of $10^{16} \, e^+$/s, is by four orders of magnitude higher than what has been achieved so far[2]. A new concept for a high-intensity positron source remains thus to be proposed for the LEMMA scheme to become feasible.

Two positron production schemes for the International Linear Collider (ILC) are being presently studied: (1) a baseline scheme which is based on passing a high-energy electron beam through a helical undulator to generate an intense photon beam for positron production in a thin target, and (2) a scheme based on the use of a separate and independent polarized electron beam to create electron–positron pairs in a thick target [10–12]. The undulator-based scheme allows producing a circularly polarized photon beam, enabling the generation of a longitudinally polarized positron beam. This scheme has been selected as the baseline option for the ILC. $10^{14} \, e^+$/s can be produced in this scheme, fulfilling the ILC requirements but missing the LEMMA-scheme goal by two or-ders of magnitude.

---

[*] armen.apyan@cern.ch
[1] The most intense muon beam is the $\mu$E4 beam line at the Paul Scherrer Institute in Switzerland which produces $I_m = 4 \times 10^8$ muons/s with the momenta of about 28 MeV/c.

[2] An overview of ever existed and future positron sources can be found e.g. in [9] and references quoted therein.

In this paper, we propose and present our initial exploratory studies of a new scheme for the high-intensity muon source. This scheme, as we show in the following, is capable to produce more than $10^{13}$ of both $\mu^+$ and $\mu^-$ per second and – as a bonus – more than $10^{16}$ polarized positrons per second. In the proposed scheme, the combined muon and electron/positron source is driven by the photon beam.

Photon beam-driven muon sources have never been studied before because no technology has been proposed to create photon beams of sufficiently high energy and power to be competitive with the proton-beam-driven sources. The recent Gamma Factory (GF) proposal [13] opens the path in this direction. The Gamma Factory can deliver the requisite technological breakthrough to achieve the requisite leap in photon-beam intensities by at least six orders of magnitude with respect to existing photon sources, allowing the creation of multi-megawatt photon beams.

The underlying idea of the Gamma Factory is to generate the photon beams by resonant back-scattering of laser photons off atomic beams of partially stripped ions (PSIs) circulating in the LHC storage rings. The energy of the photon beam can be tuned by choosing: (1) the ion beam specified in terms of its atomic (nuclear charge) number $Z$ and the number of non-stripped electrons, (2) the laser type, and (3) the energy of the LHC beam. The maximal photon beam energy to be achieved is $\sim 400$ MeV for the present magnetic-field limit of the LHC dipoles. It can be extended to $\sim 1.6$ GeV if the HE-LHC upgrade [14] is realised in the future. For more details on the Gamma Factory project and its possible applications see Ref. [15].

At first sight, photons have an apparent disadvantage with respect to protons. The cross-section of their collisions with nuclei producing muons in the final state – either directly or via produced-pion decays – is two orders of magnitude lower than the corresponding cross section for protons. As a consequence, the number of the produced muons per one beam particle will always be smaller for photons than for protons. Therefore, to be competitive, this disadvantage must be compensated by the higher-megawatt power of the photon beams[3] and/or by the lower cost of the photon-driven source, and/or by the higher efficiency of forming *low-emittance* muon beams.

Photon-beam-driven sources also have several unique advantages that may become important for designing high-precision experiments using muons and neutrinos coming from such sources, which the proton-beam-driven ones do not have: (i) the photon beams assure the exact charge symmetry in the spectra of produced muons, if produced by photon-conversions into $\mu^+\mu^-$-pairs, and the quasi-exact charge symmetry, if generated by decays of pions produced in photo-excitation of the $\Delta$-resonances in isoscalar nuclei targets, (ii) circularly polarized photon beams produce longitudinally polarized muon beams, (iii) single pions produced in the exclusive photo-excitation of the $\Delta$-resonances are quasi-monochromatic – their decays produce muon flux which is characterised by a sharp correlation of the muon energy and its production angle, (iv) such a sharp correlation opens the possibility to control the polarization of the collected muons, (v) if the initial muon polarization is preserved over the muon beam acceleration and storage phase, the flavour content of the neutrino and antineutrino beams generated by the polarized muon beams can be tuned and experimentally controlled with high precision.

The above potential merits of a photon-beam-driven muon source call for detailed studies of its concrete GF-based realisation. While detailed simulations and experimental studies of the proton-beam-driven scheme have been made over the last 50 years, the photon-beam-driven scheme has never been seriously studied. The studies presented here can thus be considered as the initial exploratory step going in this direction. This paper summarises the results of this initial step.

In Section II, we present our evaluation of the following two possible schemes to produce the muon beams: (1) a low-intensity scheme based on direct photon conversion into muon pairs in the electromagnetic (EM) fields of target nuclei and (2) a novel, high-intensity scheme – proposed for the first time in this paper – based on single-pion production by the resonant photo-excitation of the $\Delta$-resonances in the nuclear targets.

In the subsequent sections of this paper, we focus our attention solely on the latter, high-intensity scheme, leaving the former one to a separate report.

Section III summarises the potential advantages of the proposed photon-beam-driven muon production scheme and discusses its complementarity with respect to the proton-driven scheme.

In Section IV, we present our optimisation studies of the photon-beam target parameters. Their goal is to maximise the pion-production rate and their spectral density. We subsequently discuss the constraints coming from the tolerable photon-beam energy dissipation in the target material. This is followed by the presentation of a scheme to select pions and positrons coming out from the target. We finalise this section by comparing the photon-beam-driven pion production scheme with the proton-beam-driven one. The discussion presented in this section is restricted to a monochromatic, point-like photon beam in order to factorise out, in the initial step, the driver-beam optimisation aspects from the target optimisation aspects.

In the subsequent step, presented in Section V, we go beyond the monochromatic photon beam case. We propose and discuss the underlying principles which allow the generation of the GF megawatt-power photon beams and, subsequently, we put forward three concrete scenarios of their realisation. Presentations of the simulation tools and simulation results for one of the photon-beam scenarios close the discussion of the GF photon-beam optimisation aspects.

---

[3] Photons are *neutral* bosons while protons are *charged* fermions, therefore it is easier to produce multi-megawatt-range high-intensity photon beams than proton beams of the same power.

Having optimised both the photon beam and the target aspects of the GF muon and positron source, we present in Section VI its principal characteristics, including the expected pion, muon and positron fluxes. Section VII contains a short discussion of the necessary future steps on the path to the technical realisation of the photon-beam-driven muon and positron source. Finally, Section VIII presents conclusions of the initial phase of our studies.

## II. PRODUCTION OF MUONS BY PHOTONS

There are two ways to produce muons with the photon beam. The first one is to convert photons directly into muon pairs in the EM field of the target nuclei. The second one is to produce pions in the collisions of photons with the target nuclei, and subsequently to collect muons coming from decays of the pions. Both these schemes are discussed below.

### A. Photon conversion

The reaction discussed in this section is the quasi-elastic conversion of the photon in the electromagnetic field of the target nucleus, characterised by its mass number $A$:

$$\gamma + A \to \mu^+ + \mu^- + X \,. \tag{1}$$

Conversions of high-energy, circularly-polarized photons produce longitudinally polarized leptons, see e.g. [16]. The Gamma Factory photon source can be configured to produce circularly polarized $\gamma$-rays by using spin-0 helium-like PSI beams circulating in the LHC as the frequency converters of circularly polarized laser photons [13].

The minimal photon energy to produce a muon pair in a collision with a stationary heavy nucleus is $E_\gamma^{\min} = 2m_\mu = 211\,\mathrm{MeV}$, neglecting the nucleus recoil momentum. Photons convert mainly to $e^+e^-$ pairs. At asymptotically large photon energies, $E_\gamma \gg E_\gamma^{\min}$, photons convert to electron–positron pairs by a factor of $m_\mu^2/m_e^2 = 3.7 \times 10^4$ more often than to muon pairs. The maximal energy of the GF photon beam which can be reached with the present LHC configuration is $\sim 400\,\mathrm{MeV}$, and $\sim 1600\,\mathrm{MeV}$ if the HE-LHC upgrade option is realised [17].

Since the above energies are close to the muon-pair production threshold, mass-threshold effects have to be taken into account while calculating the muon-pair production cross section[4]. The reduction of a muon-pair production cross-section, with respect to its high-energy asymptotic value, while approaching the muon-pair production threshold energy is illustrated in Fig. 1 for a graphite target. The cross-section diminishes $\sim 60$ times when the photon-beam energy changes from $1600\,\mathrm{MeV}$ to $400\,\mathrm{MeV}$.

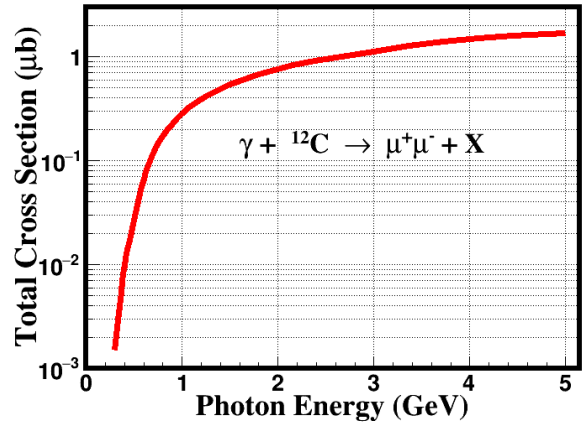

FIG. 1: The muon-pair production cross section as a function of the photon energy for a carbon target.

The calculation of the muon-pair production rates for photons colliding with long targets has to take into account the realistic lateral evolution of the photon spectra at various depths of the target. Our simulations using the GEANT4 package [18] include this effect. The results of these simulations can be summarised as follows. For a $2X_0$ long beryllium target, the number of produced muon pairs per one incoming photon increases from $2.2 \times 10^{-8}$ for $E_\gamma = 400\,\mathrm{MeV}$ to $6.4 \times 10^{-7}$ for $E_\gamma = 1\,\mathrm{GeV}$[5]. The muon-pair production rate does not vary very much with the change of the atomic number $Z$ of the target nucleus for the same target length expressed in the units of the $Z$-dependent radiation length. For example, at $E_\gamma = 800\,\mathrm{MeV}$, it decreases from the value of $3.9 \times 10^{-7}$ for the $2X_0$ graphite target to the value of $3.1 \times 10^{-7}$ for a $2X_0$ tungsten target.

These studies show that for photon beams of a megawatt power, the expected muon-production rates are comparable to those of the presently existing muon sources. The photon-conversion scheme can extend the functionality of these sources by producing longitudinally polarized muons and by assuring their perfect charge symmetry.

---

[4] One of the authors (MWK) is indebted to K.T. McDonald to for drawing our attention to the importance of the energy threshold effects. Following his comments, these effects were implemented in the new version of the GEANT4 code [18]. This version was then used in our studies for simulations of photon-beam collisions with target nuclei.

[5] This value is, by more than one order of magnitude, lower than its high energy asymptotic value.

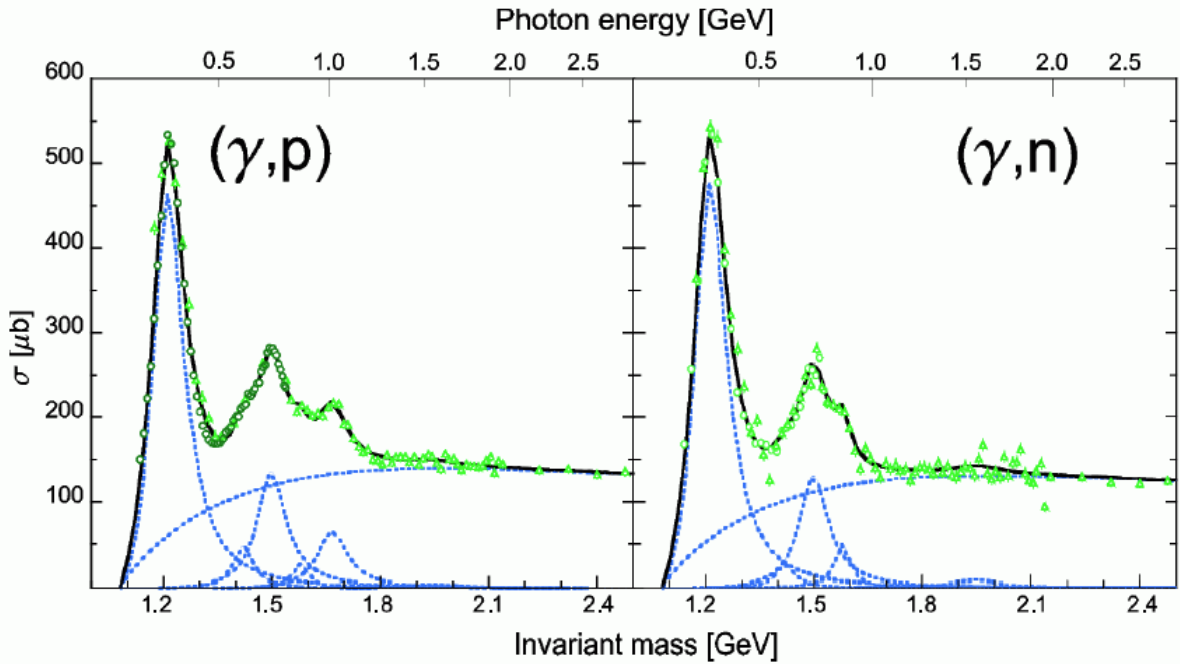

FIG. 2: Total cross sections [19] for photo-absorption on the proton (left) and the neutron (right) targets. Points: measured data, curves: fits of the Breit–Wigner shapes of nucleon resonances.

### B. Photo-production of pions

An alternative scheme for a high-intensity muon source, proposed for the first time and studied in detail in this paper, is based on the resonant photo-production process:

$$\gamma + A \rightarrow \pi + X, \quad (2)$$

followed by the $\pi \rightarrow \mu + \nu$ decays.

For the available energy range of the Gamma Factory photons, the processes involving the production of strange and charm particles are kinematically forbidden. Therefore, the GF muons are produced essentially by pion decays[6].

The measured photo-absorption cross-sections for proton and neutron targets are shown in Fig. 2 as a function of the photon beam energy or, equivalently, as a function of the collision centre-of-mass energy [19]. The peaks seen on these plots correspond to the production of the $\Delta(1232, 1600, 1620)$ and $N^*(1532, 1875, 1900)$ resonances. The GF photon energy can be tuned to excite only the lowest mass, albeit the highest cross section, $\Delta(1232)$ resonance which decays into a single pion and a nucleon. The peak cross section for the photo-production of this resonance, $\sim 0.5$ mb, is sufficiently high to produce of the order of $10^{13}$–$10^{14}$ pions per second by colliding an energy-tuned 1 MW GF photon beam with stationary nuclear targets containing protons and neutrons.

### C. Exclusive pion production in $\gamma$–nucleon and $\gamma$–nucleus collisions

The wavelength of photons that carry sufficient energy to excite the $\Delta(1232)$ resonance is smaller than the size of the nucleus. Therefore pions produced in $\gamma$–nucleus collisions originate predominantly from the binary $\gamma$–proton and $\gamma$–neutron processes:

$$\gamma + p \rightarrow \pi^+ + n, \quad (3)$$

and

$$\gamma + n \rightarrow \pi^- + p, \quad (4)$$

with the remaining nucleons acting as spectators.

To maximise the symmetry (equality) for positive and negative pion production spectra, the photon beam target must be made of material characterised by equal numbers of protons and neutrons. Isoscalar nuclei, for which $A = 2 \times Z$, satisfy this requirement.

Protons and neutrons confined in the nuclei are not stationary but move with the Fermi momenta. For a precise tuning of the photon beam energy to excite the pion-production resonances and for maximising the spectral density of produced pions, $d^2N/(dp\, dp_T^2)$, Fermi motion (FM) effects have to be taken into account and precisely evaluated. In order to understand their impact, we start our discussion with the simplest case of the collisions of photons with stationary protons in a hydrogen target.

---

[6] The direct muon-pair processes, discussed in the previous subsection, represent a small contribution and are neglected in the studies presented in this paper. As an example, for the $2X_0$ beryllium target and the photon energy below 400 MeV, this contribution stays always below 0.05%.

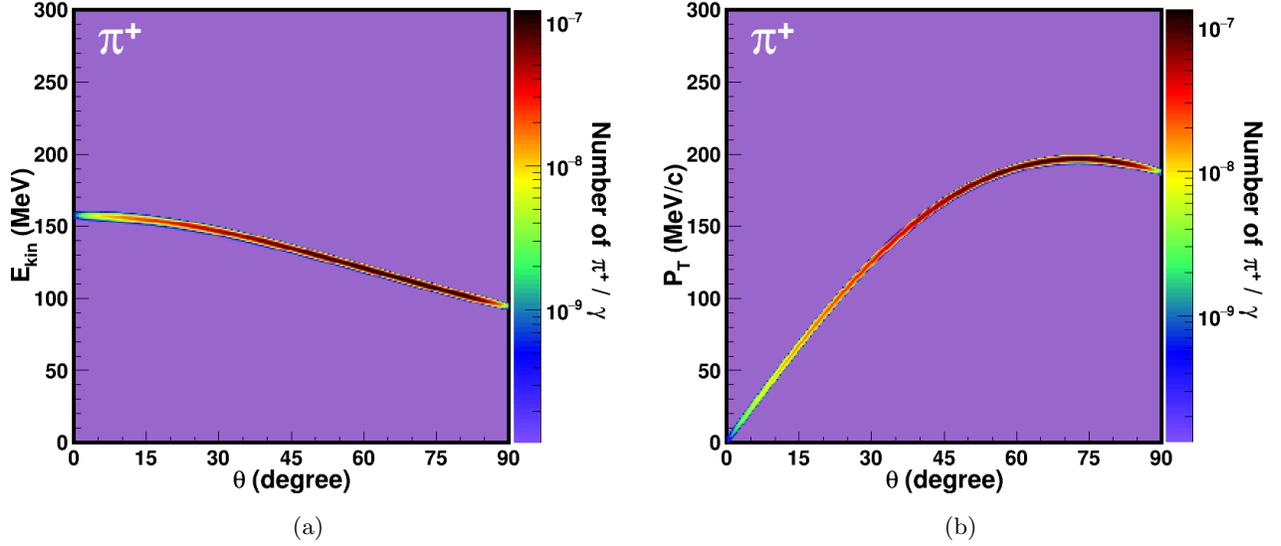

FIG. 3: Correlations of the kinetic energy (a) and the transverse momentum (b) of pions with their polar angles in the laboratory frame for a photon-beam energy of 300 MeV colliding with a hydrogen target. The differential pion rates, per incoming photon, correspond to a transverse size of the target of 2 mm and a length of 8 cm.

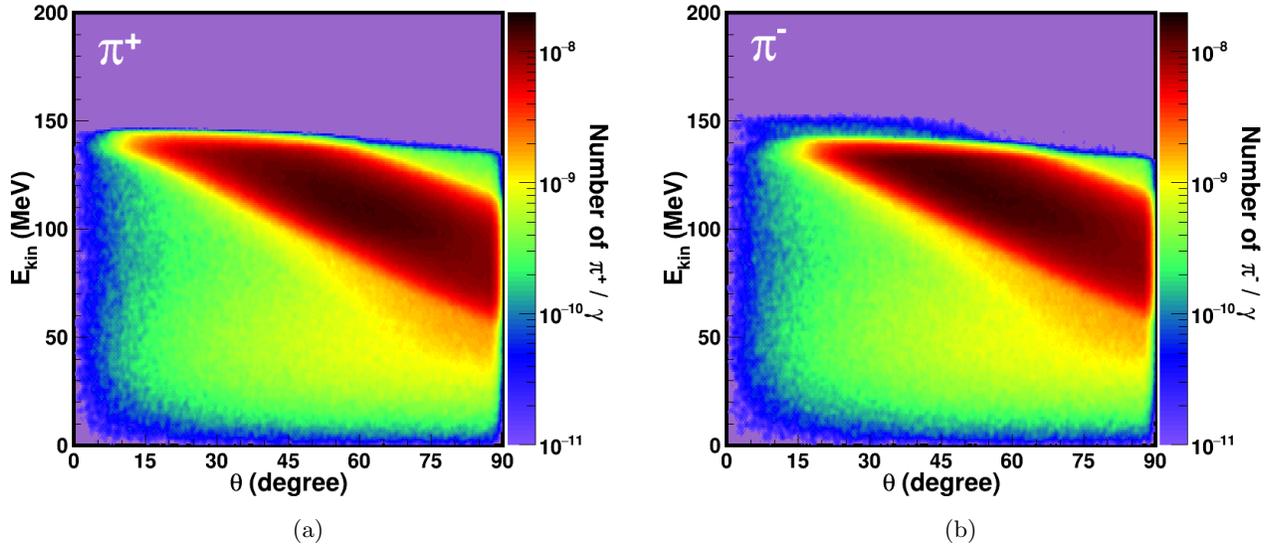

FIG. 4: Correlations of the kinetic energy with the production angle for positively (a) and negatively (b) charged pions produced by a 300 MeV photon beam colliding with a graphite target. The differential pion rates, per incoming photon, correspond to a target with a transverse size of 2 mm and a length of 2 cm.

### D. Hydrogen target

The pion production reaction (3) is a $2 \to 2$ process. For such a process, the momentum/kinetic energy of the produced pion is, in the collision centre-of-mass frame, no longer random but fixed (for a stationary target and for a fixed photon-beam energy). Therefore, in this reference frame, photon–proton collisions produce monochromatic positively charged pions. The stochastic aspect of this process is restricted to the direction of the outgoing pion, specific to decays of the spin 3/2 $\Delta$-resonances. In Fig. 3, correlations of the kinetic energy (a) and the transverse momentum (b) of the produced pions with their polar emission angles in the laboratory frame are shown for a 300 MeV photon beam colliding with a hydrogen target. These plots show that the Lorentz transformation from the centre-of-mass to the laboratory frame smears the pion kinetic energy, albeit in a deterministic manner.

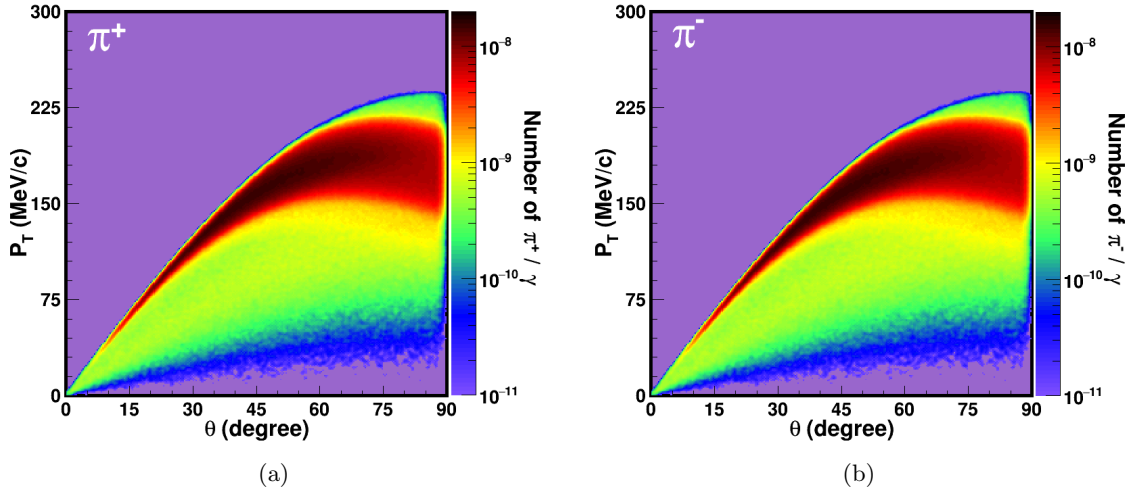

FIG. 5: Correlations of the transverse momentum with the production angle for positively (a) and negatively (b) charged pions produced by a 300 MeV photon beam colliding with a graphite target. The differential pion rates, per incoming photon, correspond to a target with a transverse size of 2 mm and a length of 2 cm.

The pion kinetic energy (as well as its transverse momentum) becomes a function of a single parameter – the pion emission polar angle.

This *exclusive pion production process* opens thus a path to design a pion collection scheme which can maximise the spectral density of the collected pions. For a pion collection scheme in which the kinetic energy of the produced pions is "corrected" as a function of the pion production angle, and the momentum vector is rotated by a suitably chosen toroidal magnetic-field configuration to reduce its transverse component, the emittance of the pion source can approach that of the driver photon beam.

Protons in the hydrogen target are stationary, while nucleons in nuclear targets are not. Their Fermi motion, as discussed in Subsection II E, is the principal, irreducible source of deterioration of the spectral density of pions produced with nuclear targets.

### E. Isoscalar targets

In Fig. 4 and Fig. 5, the correlations of the kinetic energy and the transverse momenta of the produced positively and negatively charged pions with their polar emission angles are shown for a 300 MeV photon-beam colliding with a graphite target.

These plots illustrate the quasi-perfect symmetry of the positive and negative pion spectra for the collisions of a photon beam with an isoscalar target. In addition, they illustrate the effect of the smearing of the pion kinetic energy and transverse momentum, caused by FM of protons and neutrons.

This FM effect is quantified in Fig. 6 where the pion kinetic energy and transverse momentum are shown, together with their smearing, as a function of its polar emission angle. The error bars on these plots have a special meaning. They represent the standard deviation $\sigma_{\text{FM}}(\theta)$ of the kinetic energy (transverse momentum) distribution in the respective $\theta$ bins. The FM smearing of the pion kinetic energy (transverse momentum) varies in the range of 10 MeV for pions produced at small angles $\theta$ to 30 MeV for pions produced perpendicularly to the photon–nucleus collision axis.

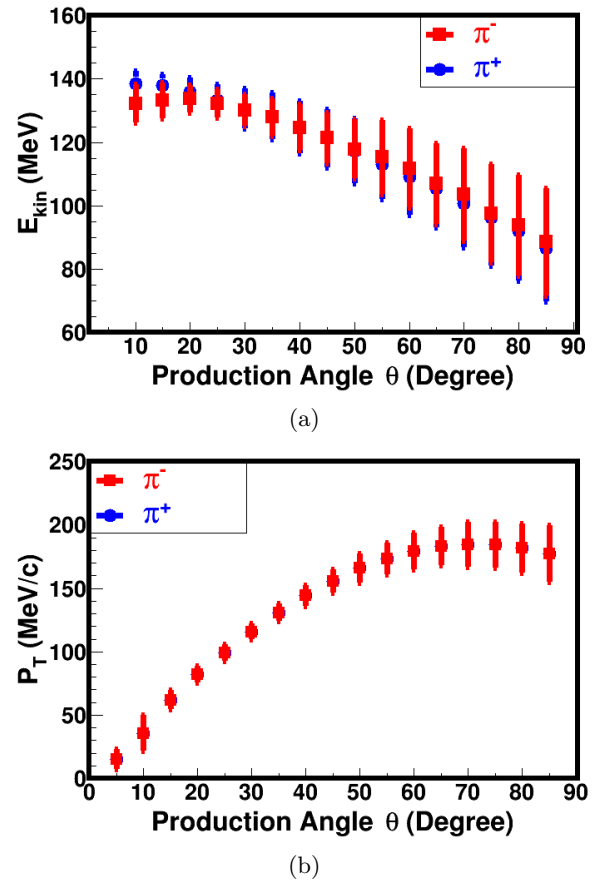

FIG. 6: The pion kinetic energy (a) and transverse momentum (b), together with the standard deviation of their smearing represented by the error bars, as a function of the pion-production angle for the graphite target.

The $\theta$-dependent average correction of the pion kinetic energy and transverse momentum as a function of the pion emission angle – allowing to maximise the spectral density of pions – can, in principle, be made by a suitable design of the target and the pion-collection system. The FM smearing cannot, however, be "corrected for" and will determine the irreducible "stochastic" contribution to the pion source emittance.

## III. MERITS OF PHOTON-BEAM-DRIVEN MUON SOURCE

### A. Charge symmetry

Proton beams produce significantly more positively than negatively charged pions. This asymmetry is particularly strong for low-momentum proton beams [20] for which kaon production is kinematically forbidden. For a proton-beam energy in the range of 290 – 600 MeV, where only single-pion production processes are kinematically allowed, all the existing muon sources produce positively charged muons only. The charge symmetry of the produced positive and negative pions is gradually restored at high proton-beam energy. However, even at a proton beam energy of 8 GeV, the ratio of $\pi^+/\pi^-$ yields remains to be above a factor of 1.5. The price to pay for partially restoring this charge symmetry is to expand the kinematical region of the produced pions by using high-energy proton-driver beams producing both pions and kaons. Forming low-emittance muon beams coming from such a pion source is anything but easy. Moreover, the fluxes of neutrinos coming from the broad momentum-spread kaon and pion mixed sources have never been controlled with a percent-level precision.

A photon-beam-driven source, based on the exclusive single-pion production is expected, on the contrary, to be charge symmetric for the beam energy tuned below the two-pion production threshold and using isoscalar targets[7]. A muon beam created by such a source is expected to be quasi charge-symmetric. Such a beam is optimal for precision studies of neutrino/antineutrino CP asymmetries.

### B. Suppression of kaon decays

Proton beams – in the energy range where the charge symmetry of the produced mesons is partially restored – produce both pions and kaons. Kinematical distribu-tions of the muons and neutrinos coming from pion and kaon decays are different. They reflect both the differences in their respective decay modes and their different masses. Precise control of the muon and neutrino fluxes will always be limited by the uncertainty of the relative yields of pions and kaons which cannot be controlled by QCD calculations and has to be measured. Moreover, the average transverse momentum of muons energy coming from the principal decay mode of kaons $K \to \mu\nu$ is $P_T^\mu(K) \sim (m_K - m_\mu)c/2$, and its spread is significantly larger than that for pion decays $P_T^\mu(\pi) \sim (m_\pi - m_\mu)c/2$. The presence of kaons in the beams that produce muons and neutrinos increases the emittance of muon sources and gives rise to a two-fold uncertainty of the neutrino momentum at its fixed emission angle.

---

[7] For isoscalar nuclei, the remaining small asymmetries come from (1) the diffrence of the $u$ and $d$ valence-quark electric charges, (2) the proton and neutron mass difference and (3) the pion-charge-dependent final-state electromagnetic interactions [19].

A photon-beam-driven source can use a photon beam with an energy below the kaon production threshold. The kaon contamination effects, discussed above, are thus no longer of any importance for such a source.

### C. Spectral density

High-energy protons colliding with stationary targets produce pions and kaons over a wide momentum range. As a consequence, the emittance of the proton-beam-driven pion/kaon source and, subsequently, the muon source is large and must be reduced by several orders of magnitude to form a beam that can be efficiently accelerated. At present, ionisation cooling [21] is the only available option to reduce the initial large emittance of a proton-beam-driven muon source. The principle of the ionisation cooling was recently demonstrated at a single-particle level by the MICE experiment [22]. It still remains to be demonstrated for the case of high-intensity muon beams.

The principal goal of the photon-driver beam scheme, proposed in this paper, is to reduce both the momentum and the transverse-momentum phase-space regions of the produced pions in order to maximise their spectral density to a level which is limited only by the irreducible target nucleon Fermi-motion smearing, as discussed in Subsection II E. Pions produced in this scheme are quasi-monochromatic. Their spectral density is expected to approach that of the proton-beam driven scheme for the same driver-beam power – even if the cross-section for producing muons by photons is a factor of up to 100 smaller than that for protons.

### D. Narrow-band neutrino beams

Monochromatic pion and kaon beams produce neutrino narrow-band beams (NBBs). For such beams, the energy of the neutrino is determined by its angle reconstructed in a neutrino detector. For a mixed pion/kaon beam, a di-chromatic neutrino beam is formed, exhibiting a characteristic two-fold ambiguity of the neutrino energy as a function of the fixed value of its reconstructed angle. Di-chromatic neutrino NBBs have always played an important role in measurements requiring precise control of neutrino fluxes, for example in the measurements of the energy dependence of the neutrino cross sections. In proton-beam-driven neutrino sources, pions and kaons are produced initially with a large momentum spread. Therefore, neutrino NBBs are produced by selecting only the small fraction of pions and kaons produced in a narrow momentum region at the expense of a significant reduction of neutrino-beam intensity.

A photon-beam-based scheme can be optimised to reach the highest achievable pion spectral density already at the pion-production phase, such that no further momentum selection would be necessary. Moreover, the two-fold ambiguity of the neutrino energy inherent to the dichromatic beams is avoided by getting rid of the kaon production processes.

### E. Muon polarization

In the rest frame of a decaying pion, the spin of the outgoing muon is aligned with its momentum vector. The transformation to the laboratory frame involves the Lorentz boost and the Wigner rotation of the muon spin. The polarization of each of the muons in the laboratory frame is thus fully specified by the parent-pion momentum and by the pion-decay polar angle in this reference frame.

In order to form beams of highly polarized muons, the pion source must produce monochromatic pions[8], and the muon collection system must keep track of the muon emission angle, e.g. by collecting the muons emitted in narrow decay-angle $\theta$-slices in separate rf-buckets. The first of these two conditions cannot be fulfilled for high-energy proton-beam-driven sources due to the large momentum spread of the produced pions. In addition, for the presently considered collection schemes of the ionisation-cooled muons based on solenoidal field beam-focusing, the information on the initial muon emission angle is lost in the muon collection and cooling phase. Therefore, these sources cannot produce beams of highly polarized muons.

The photon-beam-driven source produces quasi-monochromatic pions fulfilling the first of the two necessary conditions to form the polarized muon beam, and opens the possibility of designing the muon collection and acceleration scheme which preserves the information on the initial muon polarization while forming beams of polarized muons.

### F. CP and flavour composition of neutrino beams

The availability of high-energy beams of polarized muons can open new unprecedented research opportunities for neutrino physics.

---

[8] A special version of "monochromatic" muons source is one using a target-stopped muons. Another way to control the muon polarization – for a very low-intensity muon beam – is to tag the momentum vector of each of the decaying pions.

The two most important challenges in neutrino physics are: (1) the precise measurement of the neutrino masses and establishing their hierarchy, and (2) the measurement of the CP violating phase in the neutrino-mixing matrix. The CP violation in the neutrino sector leads to a difference of the oscillation pattern from one neutrino flavour into another neutrino flavour between neutrinos and antineutrinos.

From the experimental point of view, the comparison of the oscillations of $\nu_\mu \to \nu_e$ and $\bar\nu_\mu \to \bar\nu_e$ is the most promising way to search for CP violation. This needs ideally a pure $\nu_\mu$ beam, without a $\nu_e$ background, and a pure $\bar\nu_\mu$ beam, without a $\bar\nu_e$ background, to guarantee the best sensitivity for the appearance of $\nu_e$ and $\bar\nu_e$, respectively. An additional important requirement is the precise relative normalisation of the initial $\nu_\mu$ and $\bar\nu_\mu$ fluxes.

For fully-polarized muon beams, the flux of muon neutrinos produced in the direction of the incoming beam is enhanced, while the electron neutrino flux is zero. This is the most favourable configuration to search for CP violation in neutrino oscillations. The experiment would measure and compare the oscillation probabilities of $\nu_\mu \to \nu_e$ with the $\nu_\mu$ flux obtained from the decay at the zero polar angle from the fully polarized $\mu^-$, and of $\bar\nu_\mu \to \bar\nu_e$ with the $\bar\nu_\mu$ flux obtained from the decay at the zero polar angle from the fully polarized $\mu^+$. An ideal experiment would require (separately) neutrinos from the $\mu^-$ beam (with $P_\mu = +1$) and antineutrinos from the $\mu^+$ beam (with $P_\mu = -1$), where $P_\mu$ is the muon polarization along the muon beam direction.

The photon-beam-driven scheme, discussed in this paper, cannot produce fully-polarized $P_\mu = \pm 1$ muon beams. However, it has the potential to provide beams of well-controlled and tunable polarizations, allowing for the precise control of the residual admixture of $\nu_e$ in the $\nu_\mu$ beam and $\bar\nu_e$ in the $\bar\nu_\mu$ beam.

All the above merits call for detailed studies of the photon-beam-driven muon source. Its initial, exploratory, stage is summarised in the rest of this paper.

## IV. TARGET OPTIMISATION STUDIES

The initial step in the exploratory studies of a photon-beam-driven muon source is the optimisation of the photon beam target. The principal goal of this step is to maximise the number of pions produced and their spectral density.

The studies presented in this section are made for a simple case of a monochromatic, point-like, zero-divergence, 1 MW ($N_\gamma = 2 \times 10^{16}$ photons/s) beam of photons with an energy of 300 MeV.

The pion production cross section in photon–carbon collisions is shown in Fig. 7 as a function of the photon energy. Our choice of 300 MeV maximises the pion production cross section, minimises the charge asymmetry of the produced pions and suppresses the contribution of processes leading to the production of more than one pion. Extension of these studies to a more realistic

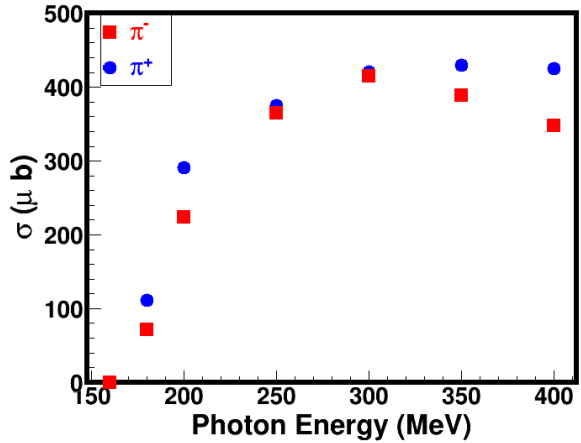

FIG. 7: The pion-production cross section in the photon–carbon collisions as a function of the photon energy for pions produced in the forward region ($\theta \leq 90°$).

Gamma Factory photon-beam is presented in Section VI.

### A. Material

The number of pions produced depends on the material type and the geometrical dimensions of the target. Monte Carlo simulations using GEANT4 [18] were performed to study the photoproduction of pions for several materials spanning the full range of the atomic number $Z$. Definitions of deuterium, graphite and calcium targets, used in the simulations, are taken from the standard GEANT4 material database [18].

The maximal rates of positive and negative pions produced can be obtained by using deuterium targets. For a $2X_0$-long deuterium target range, the rate of the produced positive and negative pions is $2.1 \times 10^{14}\,\mathrm{s}^{-1}$. For targets with a fixed length of $2X_0$, the pion rate decreases with an increasing value of the atomic number $Z$ of the target material, and their charge asymmetry increases.

As an example, the kinetic-energy-dependent production rate of positive and negative pions in the forward direction due to a 300 MeV, 1 MW photon beam, colliding with a cylindrically-shaped, 1 cm radius, 40 cm long graphite and 20 cm long calcium targets[9] are shown in Fig. 8.

The GEANT4 simulations show a small asymmetry in the $\pi^-$ and $\pi^+$ spectra for all the isoscalar target materials. This charge asymmetry was found to be independent of the target length and radius and to be uniquely driven by the present GEANT4 model for pion production in photon–nucleus collisions. The model implemented in the GEANT4 code includes the charge-dependent final-state interactions of the produced pions with the spectator nucleons of the target nucleus. The uncertainty of this model is large (comparable to the size of the asymmetries) and its predictions need to be verified by dedicated measurements.

---

[9] Both targets, when expressed in radiation-length units, are approximately $2X_0$-long.

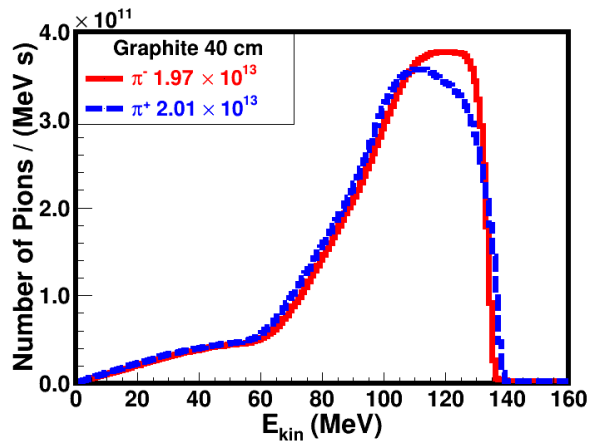

(a)

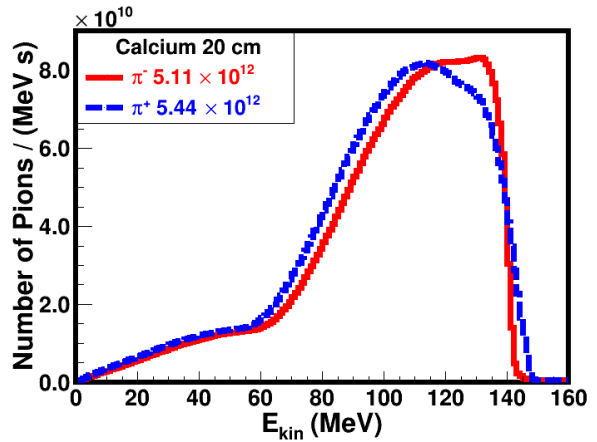

(b)

FIG. 8: Pion rates and kinetic energy distributions for the 300 MeV, 1 MW photon beam colliding with the 40 cm long ($2.07X_0$) graphite target (a) and the 20 cm long ($1.92X_0$) calcium target (b), both of the cylindrical shape with the 1 cm radius. Only forward-going pions are included in this plot.

The lowest-$Z$ material (deuterium) choice would maximise the pion production rate. However, the corresponding requisite length of the $2X_0$-long liquid deuterium target (of $\sim 15\,\mathrm{m}$) would be comparable to the produced-pions decay length in the laboratory reference frame. A large fraction of pions would decay while propagating through the target prior to forming a low-emittance pion beam. Graphite represents, in our view, the optimal material choice, assuring the maximal production-rate for a reasonably-small length of the pion-production zone, and it has been chosen as a production-target material for our studies presented in the following.

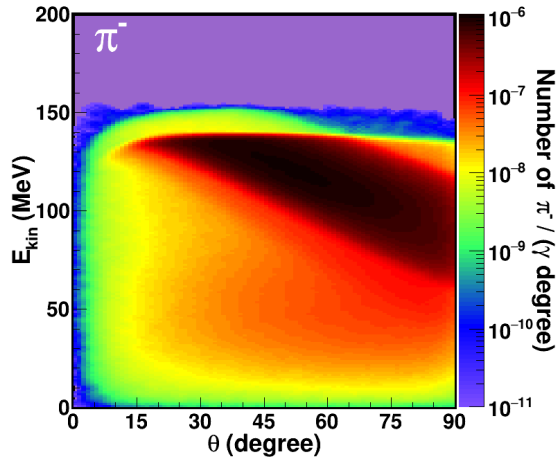

FIG. 9: Correlations of the kinetic energy and the production angle for $\pi^-$ produced in the graphite target with the length of 40 cm and the radius of 0.5 cm.

## B. Radius

The target radius should be large enough to match the transverse size of a realistic photon beam. Its optimisation involves the use of the energy-loss mechanism to reduce the RMS width of the kinetic energies of the pions coming out from the target. The optimisation procedure is explained below.

Correlations of the kinetic energy and the production angle of $\pi^-$ for a graphite target of 40 cm length and 0.5 cm radius are presented in Fig. 9. This plot shows that high-energy pions are produced at small angles, while the low-energy ones are at large angles. The mean energy loss of pions leaving the side wall of the cylinder is determined by the target radius and the pion emission angle. Fast pions produced at small angles lose more energy, on average, than pions produced at large angles before exiting the target. This is illustrated in Fig. 10: for two pions produced at $\theta_2 < \theta_1$, their corresponding path-lengths in the target are $l_2 > l_1$[10].

In order to find the optimal target radius which minimises the dispersion of the kinetic energy distribution of pions coming out of the target, we have studied targets with radii from 0.5 to 10 cm. The kinetic energy distributions and the number of $\pi^-$ produced for the varying radius of the target are presented in Fig. 11, and their mean values and standard deviation are collected in Table I. The optimal graphite-target radius which minimises the dispersion of the kinetic energies of the pions and maximises the pion-source spectral density is within the range of 2.5–5.0 cm. The final choice of the optimal target radius within this range, discussed in detail in Subsection IV D, will be driven by handling the dissipation of the energy of the megawatt-power beam in the target material.

---

[10] In our simulations, this simplified geometrical correlation is smeared by the realistic distribution of the pion-production point along the target which includes the full simulation of the EM cascade.

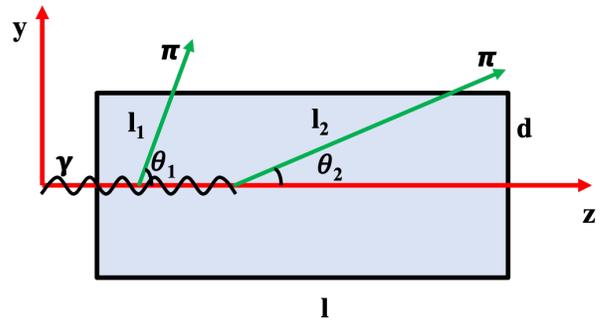

FIG. 10: Dependence of the produced-pion path inside the target on the production angle.

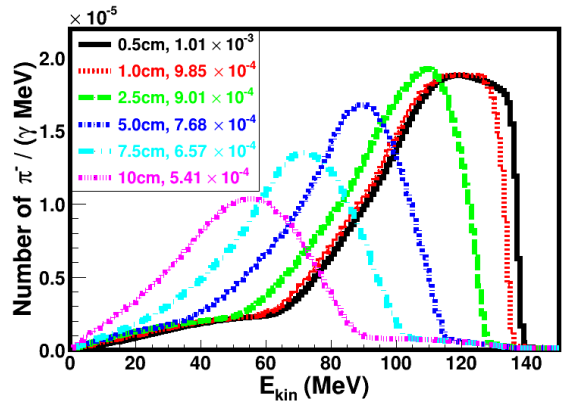

FIG. 11: Kinetic energy distributions of the $\pi^-$ produced in a $2X_0$-long graphite targets of different radii. Only forward-going pions are included in the plot. The energy of the primary photon beam is 300 MeV.

## C. Length

The pion production rates for a graphite target of variable length and the fixed radius of 2.5 cm are presented

TABLE I: The mean values and the standard deviations of the $\pi^-$ kinetic energy distributions for the varying target radius with a target length of 40 cm.

| Radius (cm) | $E_{\mathrm{kin}}^{\mathrm{mean}}$ (MeV) | $\sigma_{E_{\mathrm{kin}}}$ (MeV) |
|:---:|:---:|:---:|
| 0.5 | 116.43 | 23.92 |
| 1.0 | 114.55 | 23.70 |
| 2.5 | 102.50 | 16.13 |
| 5.0 | 85.67 | 16.65 |
| 7.5 | 69.52 | 18.60 |
| 10.0 | 52.38 | 20.86 |

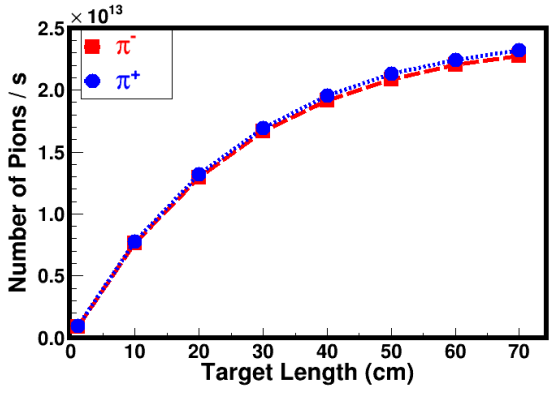

FIG. 12: The number of pions with a production angle $\theta_{\text{prod}} \leq 90°$ in the laboratory frame produced by a 300 MeV and 1 MW photon beam as a function of the graphite target length with a target radius of 0.5 cm.

in Fig. 12 and Table II.

The dominant process responsible for the flattening of the pion yield with increasing target length is the conversion of photons into electron–positron pairs in the EM field of the target-material atoms. For a target length above $1X_0$ (19.3 cm for graphite), the increase of the pion-production rate slows down, as shown in Fig. 12. The EM cascade, initiated by photon conversion into the electron–positron pairs, contributes to the target heating but hardly increases the number of pions produced. Therefore, the choice of the optimal target length must be preceded by studies of the energy deposition of the photon beam in the pion-production target. This is discussed in the next subsection IV D.

### D. Target heat load

A MW-range photon beam deposits a significant portion of energy in the target medium. The resulting increase in the target temperature causes thermal stress on the target. We have studied several candidates for the target material: graphite, aluminum, iron, copper, and tungsten – taking into account the critical material properties, especially their melting temperatures.

Dedicated Monte Carlo simulations using the GEANT4 package were carried out to investigate the deposited-energy distribution in the target over 1 second of beam operation (called in the following the deposited power), and to estimate the resulting temperature rise. Their results provide additional restrictions for the choice of the optimal target material, its length, and its radius.

The results presented below are for a graphite target. Graphite has a high melting temperature and the carbon nuclei are isoscalar. It is widely used as a beam dump [23]. The studies were made for a point-like photon beam with a fixed energy of 300 MeV. A cylinder with a 2.5 cm radius and a length of 40 cm was chosen as a target volume. The power of the incoming photon beam, hitting the target in its centre, was assumed to be 1 MW.

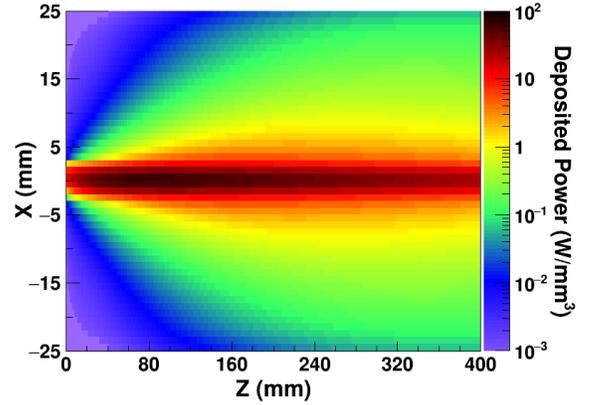

(a)

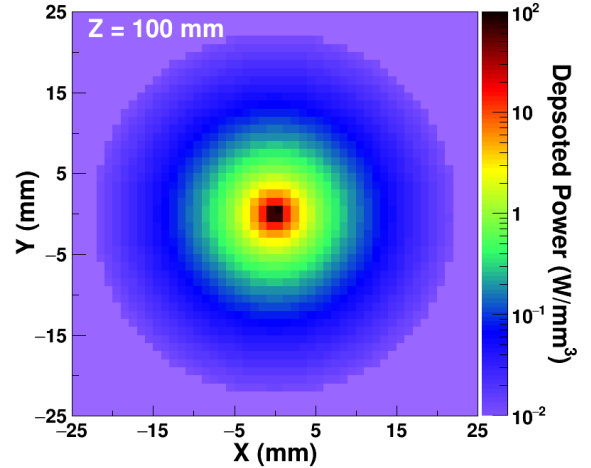

(b)

FIG. 13: Deposited beam power in the horizontal ($X$–$Z$) plane for a 1 mm thick ($Y$) slice at the target centre (a) and in the transverse ($X$–$Y$) plane for a 1 mm thick ($Z$) slice at a depth of 10 cm (b).

TABLE II: The number of pions with the production angle $\theta_{\text{prod}} < 90°$ for different lengths of the graphite target with a radius of 2.5 cm.

| Length (cm) | $N_{\pi^+} \times 10^{11}$ | $N_{\pi^-} \times 10^{11}$ |
|---|---|---|
| 1.0 | 9.28 | 9.14 |
| 10.0 | 77.64 | 76.32 |
| 20.0 | 131.78 | 129.49 |
| 30.0 | 169.42 | 166.43 |
| 40.0 | 195.32 | 191.87 |
| 50.0 | 212.91 | 208.98 |
| 60.0 | 224.42 | 220.42 |
| 70.0 | 231.93 | 227.76 |

TABLE III: The integrated beam-power deposition in a graphite target from a 1 MW, 300 MeV, point-like photon beam and its transverse width $\sigma_{x/y}$ for three values of the target length.

| Length (cm) | Energy deposition (kW) | $\sigma_{x/y}$ (mm) |
|---|---|---|
| 10.0 | 42.7 | 2.1 |
| 20.0 | 142.0 | 3.0 |
| 40.0 | 360.4 | 4.1 |

The horizontal and transverse profiles of the deposited power are shown in Fig. 13. The integrated power depositions and their transverse plane widths are presented in Table III for the three selected target lengths. The integrated power deposition and its transverse width are growing with the length of the target.

The transverse-plane-integrated deposited power for a 1 MW beam is shown in Fig. 14a as a function of the target penetration depth. It reaches its maximum value at the depth of $\sim 22$ cm, corresponding to $1.14 X_0$. Fig. 14b shows the integrated beam power deposition in the radial slices to illustrate the transverse plane diffusion of the beam power deposition with increasing beam-penetration depth.

Since the rise of the integrated deposited power with the target length is faster than the saturating rise of the pion production rate – for a target length above $1 X_0$ – our choice of the optimal graphite target length for the subsequent studies is 20 cm, corresponding to 1.03 in radiation-length $X_0$ units.

The target-volume average power dissipation in 20 cm long targets with a radius of 2.5 cm is $142 \,\mathrm{kW}/392.7 \,\mathrm{cm}^3 = 361.6 \,\mathrm{W/cm}^3$. For a forced water-cooling system, the maximum surface-heat flux that can be safely removed from the target is about $200 \,\mathrm{W/cm}^2$ [24]. This value is by a factor of two smaller than the heat flux which has to be removed from the $1 X_0$-long graphite target. A possible remedy to this would be to increase the radius of the target by a factor of two[11] or to use a target rotating in the horizontal plane.

For an illustration of the importance of the target cooling aspects, we have estimated that if the heat is not evacuated, the temperature rise in the graphite target due to the impact of the Gamma Factory 1 MW photon beam is $\Delta T \approx 230°\mathrm{C/s}$. The peak energy deposition per second (on axis) in the target can reach a value of $\sim 7 \,\mathrm{kW/cm}^3$, which corresponds to $3.2 \,\mathrm{kW/g}$ for a graphite target.

### E. Particle selection

Only a small fraction of the 300 MeV photons produce pions. The majority of them are converted in the target material to the electron–positron pairs. Electrons and positrons give rise to an electromagnetic cascade, producing secondary photons, electrons and positrons.

We propose two ways of selecting pions against the dominant contributions of electrons and positrons. The first one is to make use of the differences in their velocities $(\beta)$[12]. The second one is to exploit the difference in their production mechanism, leading to their distinct kinematical characteristics.

In Fig. 15, we present the distributions of the production angle $\theta$, in the target rest frame, for the positive and negative pions and compare them with the corresponding distributions for the electrons and positrons. Fig. 15a shows the distributions for particles reaching the propagation distance of 10 m within the time slot of $20-40$ ns following the time of the impact of the photon beam on the target, while Fig. 15b shows the corresponding distributions for the $40-60$ ns time interval. One can see a clear separation of the pions and the electrons/positrons for a photon beam which satisfies two requirements: (1) at least 50 ns time interval between consecutive photon pulses (20 MHz repetition rate) and (2) a photon pulse duration below 2 ns. The Gamma Factory photon source, discussed in Section V, fulfils this requirement.

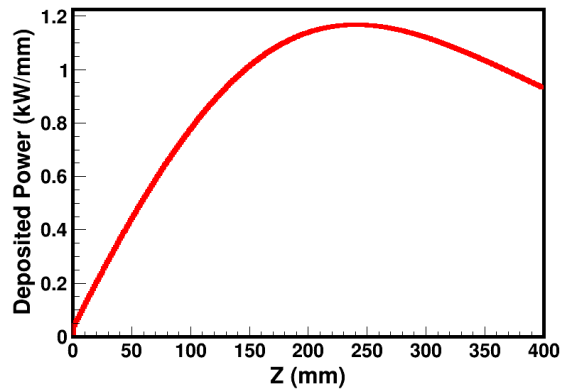

(a)

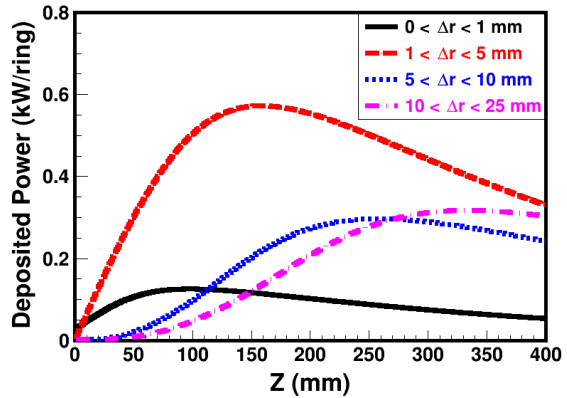

(b)

FIG. 14: The transverse-plane-integrated deposited power for the 1 MW, 300 MeV photon beam in the graphite target as a function of the longitudinal penetration depth (a), and its decomposition in the radial slices (b).

---

[11] Such an increase does not deteriorate the spectral density of the produced pions, as shown in Table I.

[12] Electrons and positrons move almost with the speed of light $(\beta_e \approx 1)$, while pions, carrying a kinetic energy comparable to their mass, are non-relativistic $(\beta_\pi \ll 1)$.

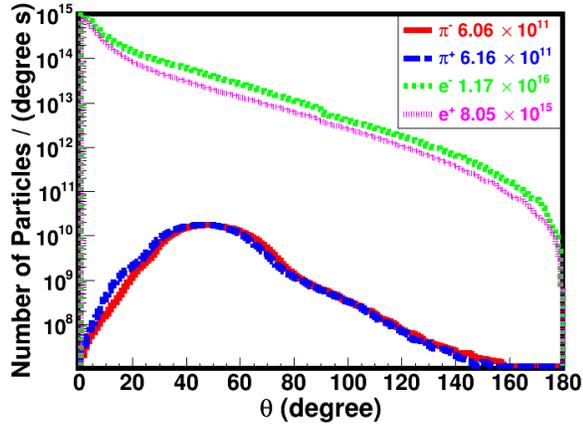
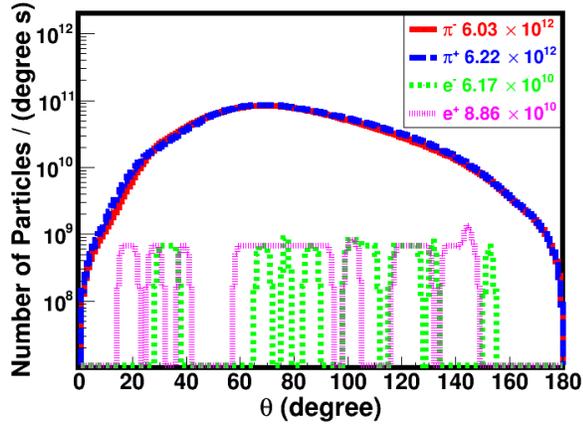

FIG. 15: Angular distributions of $\pi^-$, $\pi^+$, $e^-$ and $e^+$ at the 10 m propagation distance from the target within the time slot of $20-40$ ns (a) and $40-60$ ns (b) following the time of the impact of the photon-beam pulse on the target. The outgoing particle rates for a 1 MW photon beam are included in the figure inserts.

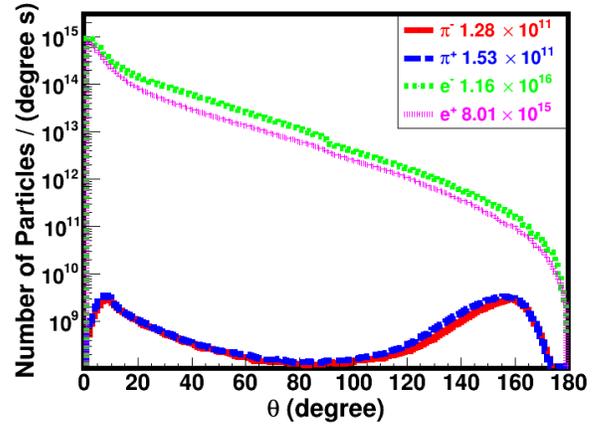
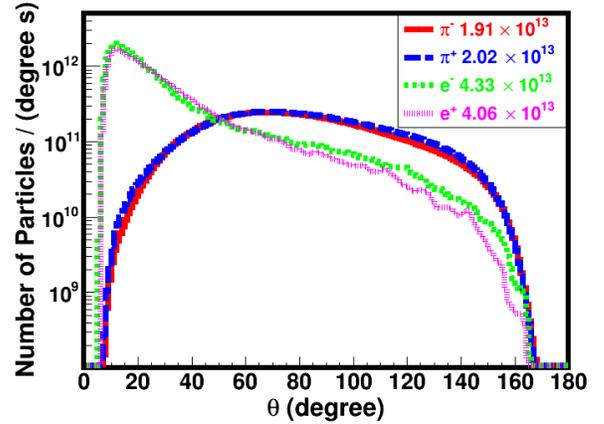

FIG. 16: Angular distributions of $\pi^-$, $\pi^+$, $e^-$ and $e^+$ at the exit of the target for the transverse-momentum cuts of $P_T < 30$ MeV/c (a) and $P_T > 30$ MeV/c (b). The outgoing particle rates for a 1 MW photon beam are included in the figure inserts.

The distinct time-of-flight characteristics of pions and electron/positrons can be used in forming pion and electron/positron beams. Pions and electrons/positrons can be collected in distinct RF-bucket trains with negligible mutual contamination.

In Fig. 16 we present again the $\theta$ distributions for positive and negative pions and compare them with the corresponding distributions for the electrons and positrons. This time, Fig. 16a includes the particles which are produced with their transverse momenta lower than 30 MeV/c, while Fig. 16b shows the particles produced with their transverse momenta higher than 30 MeV/c. The electron and positron transverse momenta are determined by the atomic form-factors. They are thus produced at small transverse momenta. On the contrary, the pion transverse momenta are determined by the nuclear/nucleon form-factors. Pions are thus produced at significantly higher transverse momenta. The distinct phase space population of pions and electrons/positrons provides a complementary way to form separately the pion and electron/positron beams[13]. Particles having transverse momenta higher than 30 MeV/c and collected in the angular region $\theta \geq 40°$ are predominantly pions. The small electron and positron contribution, in this an-gular region, can be further reduced by applying a suitable timing cut for the bunched photon beams, as discussed earlier. Particles having the transverse momenta smaller than 30 MeV/c are predominantly electrons and positrons.

---

[13] The transverse-momentum-cut separation can be achieved by introducing a solenoidal magnetic field in the target zone.

The kinetic energy distributions for electrons and positrons, and positive and negative pions are shown in Fig. 17. Electrons and positrons are selected by imposing: (1) the transverse-momentum cut $P_T \leq 30$ MeV/c and (2) the angular cut $\theta \leq 20°$. Pions are selected by imposing: (1) the transverse-momentum cut $P_T \geq 30$ MeV/c and (2) the angular cut $40° \leq \theta \leq 120°$. For the 1 MW and 300 MeV monochromatic photon beam, the number of positive (negative) pions produced over the full kinetic energy domain is $1.64 \times 10^{13}$ s$^{-1}$ ($1.57 \times 10^{13}$ s$^{-1}$). The respective numbers of electrons (positrons) produced are $8.40 \times 10^{15}$ s$^{-1}$ ($6.32 \times 10^{15}$ s$^{-1}$). The remaining small charge asymmetry in the rate of the opposite sign pions and of the electrons/positrons reflects the presence of non-leading interaction processes involving the positively charged target nuclei and the negatively charged electron cloud. In the pion case, as already discussed, it reflects the charge-dependent final state interactions of pions propagating through the nuclei. The electron/positron charge asymmetry is driven by the contribution of Compton scattering on the target electrons.

In the studies presented below, we shall use the pion and electron/positron selection method based on the transverse momenta of the produced particles.

### F. Pion spectral density

In Fig. 18, the rate of the positive and negative pions and their momentum and transverse momentum distributions are shown for the optimal target material (graphite), the optimised target length (20 cm) and ra-dius (2.5 cm), and for the selected pion-collection angular region ($40°$–$120°$).

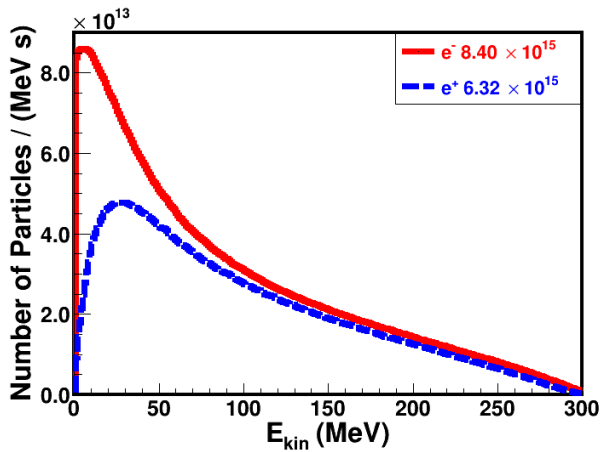

(a)

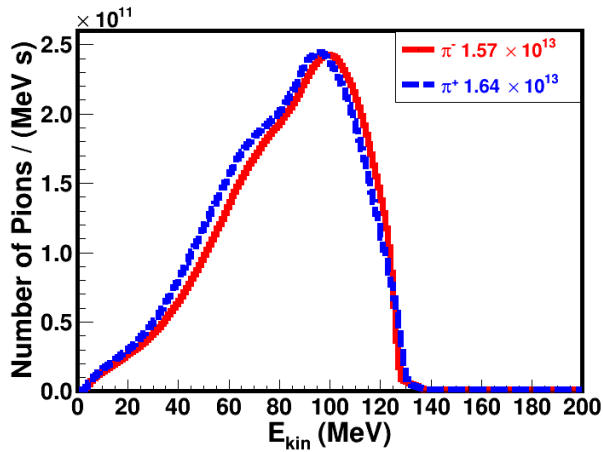

(b)

FIG. 17: Kinetic energy distributions of $e^-$ and $e^+$ (a), and $\pi^-$ and $\pi^+$ (b) at the exit of the target selected by the transverse-momentum and angular cuts specified in the text.

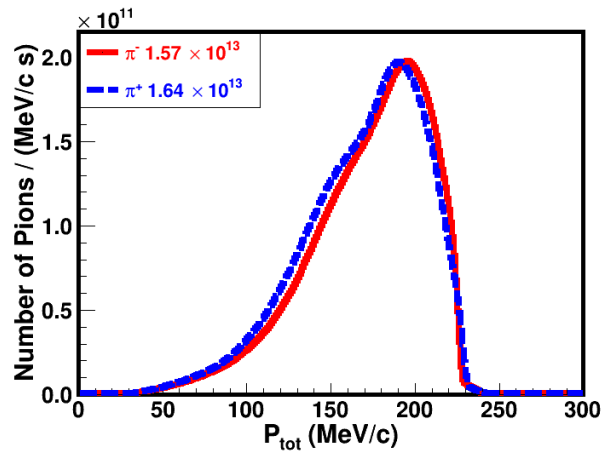

(a)

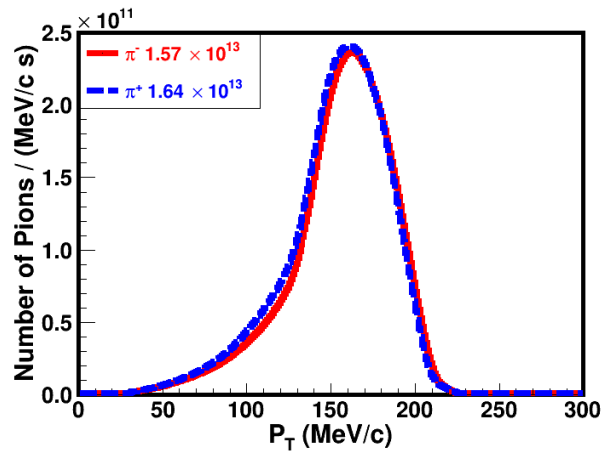

(b)

FIG. 18: The total (a) and transverse (b) momentum distributions of $\pi^-$ and $\pi^+$ at the exit of the graphite target for a 1 MW and 300 MeV photon beam with the optimised target material, target geometry and pion selection procedure (as described in the text).

This figure demonstrates that: (1) for a 1 MW and 300 MeV photon beam one can achieve pion fluxes higher than $10^{13}\,\text{s}^{-1}$ of both positive and negative pions, and (2) the pions are produced over a very narrow range of momentum and transverse-momentum distributions. The widths of the momentum and transverse-momentum distributions are, respectively, $\sigma_{P_\text{tot}} = 24\,\text{MeV/c}$ and $\sigma_{P_T} = 22\,\text{MeV/c}$.

### G. Comparison of the proton- and photon-beam-driven schemes

The proton- and photon-beam-driven muon-source intensities and emittances are determined by the characteristics of the source of the primary pions. For their unbiased comparison, we consider a 1 MW beam of photons at 300 MeV and a 1 MW a beam of protons at 8 GeV[14]. The target specification is the same for both cases: it is a graphite target with a length of 20 cm and a radius of 2.5 cm[15].

Fig. 19 shows the comparison of the pion production rates for proton and photon beams of equal 1 MW power. As expected, the proton beam of the same power produces significantly more pions. However, in any realistic scheme of pion collection, only a fraction of pions in the limited phase-space region can be used to form a muon beam. For example, if the pion collection is restricted to the momentum interval of $100\,\text{MeV/c} < P_\text{tot} < 200\,\text{MeV/c}$[16] and the transverse momentum interval of $140\,\text{MeV/c} < P_T < 180\,\text{MeV/c}$ (to reduce the spectral width and thus the initial emittance of the pion beam), then the proton beam produces only a factor of 1.8 more $\pi^-$ and a factor 2.4 more $\pi^+$ than the photon beam of equivalent power. For the above pion selection criteria, both beams can thus be considered as almost equivalent.

The exploratory studies presented in this section were performed for a hypothetical monochromatic photon beam. Their results are encouraging and call for more detailed studies based on simulations of a realistic photon beam which can be delivered by the Gamma Factory. The Gamma Factory scheme for producing a megawatt-class photon beam in the requisite energy range is discussed in the next section V.

---

[14] For the proton-beam-driven source, we have chosen the proton-beam energy of 8 GeV to follow the studies presented in [24] as well as the choice of the proton-beam energy of the Fermilab Mu2e experiment [25].
[15] For low-$Z$ materials, the radiation length and the nuclear-collision length are of comparable magnitude.
[16] The majority of pions in this momentum interval decay over the distance of 20 m from the production target, allowing the formation at such a distance of a high-purity muon beam.

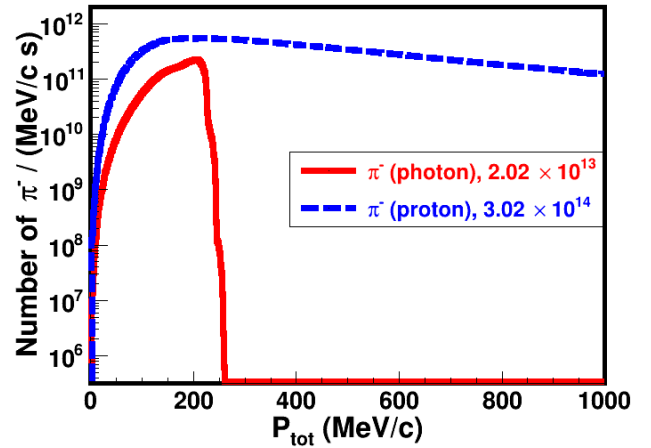

(a)

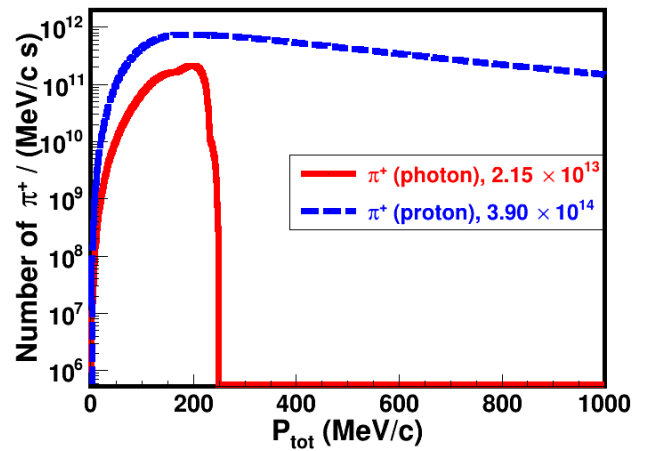

(b)

FIG. 19: The momentum distributions of $\pi^-$ (a) and $\pi^+$ (b) at the exit of a graphite target of a length of 20 cm and a radius of 2.5 cm for a 1 MW photon beam at 300 MeV and a 1 MW proton beam at 8 GeV.

## V. GAMMA FACTORY PHOTON BEAM

### A. Underlying principles

The underlying principle of the Gamma Factory photon source [13] is illustrated in Fig. 20. Highly charged, partially stripped ion (PSI) beams, with the relativistic Lorentz factor $\gamma_L$, circulate in the LHC ring. Electrons bound in these ions interact with laser photons and are excited to higher atomic energy levels. The secondary, fluorescence photons, produced in the subsequent spontaneous atomic de-excitations, are emitted in a narrow cone with an opening angle $\sim 1/\gamma_L$ in the laboratory reference frame. Their energies are boosted by a factor of up to $\approx 4\gamma_L^2$ with respect to the original laser photons used for the atomic excitation. The efficient excitation of an atomic transition relies on the resonance condition involving both the beam relativistic factor $\gamma_L$ and the laser photon frequency $\omega$. The energy band of the secondary photon beam can be varied by a suitable choice of the atomic transition (depending on the mass number $A$ of the PSI ion, its atomic number $Z$ and the number of unstripped electrons), by the corresponding tuning of $\gamma_L$ of the PSI beam and the wavelength (or the incident angle $\theta_l$) of the primary laser photons or both, to assure the resonant atomic transition:

$$E_r = \hbar \omega'_r = \hbar \omega \gamma_L (1 + \beta \cos \theta_l), \quad (5)$$

where $\omega'_r$ is a resonant photon frequency in the PSI-rest frame.

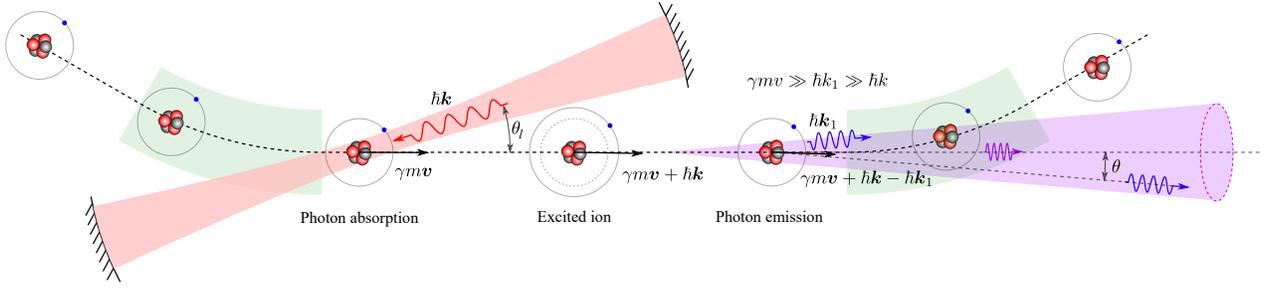

FIG. 20: The Gamma Factory concept: laser photons with momentum $k$ collide with ultrarelativistic partially stripped ions (with relativistic Lorentz factor $\gamma_L$, mass $m$, velocity $v = \beta c$, where $c$ is the velocity of light) circulating in a storage ring; resonantly scattered photons with momentum $k_1 \gg k$ are emitted in a narrow cone with an opening angle $\theta \approx 1/\gamma_L$ in the direction of motion of the ion beam.

The GF photon source is characterised by a strong correlation between the photon emission angle $\theta$ and its energy $E$:

$$E(\theta) = \frac{E_{\max}}{2\gamma_L^2 (1 - \beta \cos \theta)}, \quad (6)$$

where $E_{\max} = (1 + \beta)\gamma_L \hbar \omega'$. For small angles:

$$E(\theta) \approx \frac{E_{\max}}{1 + \gamma_L^2 \theta^2} \approx \frac{4\gamma_L^2 \hbar \omega}{1 + \gamma_L^2 \theta^2}. \quad (7)$$

This correlation can be used to form quasi-monochromatic photon beams by radial collimation of fluorescence photons. The spectral width of the collimated photon beam can be tuned by a suitable choice of: (1) the collimator slit, (2) the photon–PSI collision optics, (3) the PSI beam emittance (controlled by the transverse and longitudinal laser cooling) and (4) the distance of the secondary photon-beam radial-collimator slit to the photon production point.

The cross-section of the atomic excitation by a photon with the frequency $\omega' = \omega \gamma_L (1 + \beta \cos \theta_l)$ in the PSI rest frame can be written as [26]

$$\sigma = 2\pi^2 c r_e f_{12} g(\omega' - \omega'_r), \quad (8)$$

where $r_e$ is the classical electron radius, $f_{12}$ the oscillator strength and $g(\omega' - \omega'_r)$ is the Lorentzian function:

$$g(\omega' - \omega'_r) = \frac{1}{2\pi} \frac{\Gamma}{(\omega' - \omega'_r)^2 + \Gamma^2/4}, \quad (9)$$

where $\Gamma$ is the atomic-resonance width related to the lifetime of the excited atomic level $\tau'$ by

$$\Gamma = \frac{1}{\tau'}. \quad (10)$$

The atomic-resonance width can be expressed as

$$\Gamma = 2 r_e \omega_r'^2 f_{12} \frac{g_1}{c g_2}, \quad (11)$$

where $g_1$ and $g_2$ are the degeneracy factors of the ground and excited states, respectively.

The formula (8) can be rewritten in a simpler form:

$$\sigma(\omega' - \omega'_r) = \frac{\sigma_r}{1 + 4\tau'^2 (\omega' - \omega'_r)^2}, \quad (12)$$

where

$$\sigma_r = \frac{\lambda_r'^2 g_2}{2\pi g_1}, \quad (13)$$

and $\lambda'_r = 2\pi c/\omega'_r$ is the emitted photon wavelength. The formula in this form was used in the simulation studies presented in this section.

The polarization of the fluorescence photons can be tuned by choosing the polarization of the incident laser photons and the polarization state of the partially stripped ion. For spin-zero helium-like PSIs, the polarization of the initial laser photon can be directly transmitted to that of the secondary photon. A detailed discussion of polarization of the fluorescence photons for hydrogen- and helium-like PSIs can be found in Ref. [27].

The principal characteristics of the GF photon source can be summarised as follows: (i) *Point-like* – for high-$Z$, hydrogen- and helium-like atoms, the distance between the laser photon absorption and the fluorescence photon emission is $c \tau \gamma_L \ll 1$ cm. (ii) *High intensity* – a leap in intensity by at least 6 orders of magnitude w.r.t. the electron-beam-based Inverse Compton Sources (ICS) (at fixed $\gamma_L$ and laser pulse power). (iii) *Tuneable energy and polarization* – the tuning of the PSI beam energy, the choice of the ion, the number of left electrons and the laser type allow tuning the secondary photon-beam energy at CERN. The GF photon source, if realised, will extend by a factor of $\sim 1000$ the energy range of present FEL X-ray sources. (iv) *High plug-power efficiency* – PSIs lose a tiny fraction of their energy in the process of photon emission. There is no need to refill the driver-PSI beam. The RF power can be fully converted into the power of the secondary photon beam. (v) *Quasi-continuous beam* – the timing structure of the photon beam reflects the one of the LHC PSI beams. Photons are produced in 0.1 – 1 ns long pulses with 20 MHz frequency.

### B. Three scenarios for megawatt photon beams

In order to profit from the high cross section of the resonant pion production via $\Delta$-resonances excitation, and to assure a negligible contribution of two-pion production processes, the energy of the photon beam has to be chosen within the range of 200 – 300 MeV. To maximise the spectral density of the produced pions, the photon beam should have small energy spread – optimally, smaller than the spread of the Fermi momenta of the nucleons in the target nucleus.

The LHC operation mode to produce megawatt photon beams can follow closely its present operation mode. PSI bunches are injected from the SPS to the LHC, and following the ramp of the PSI-beam energy to its "resonant" value, are stored and exposed to collisions with laser-photon pulses. The final stripping of the electrons to the required PSI charge state is made in the SPS-to-LHC transfer line. Dedicated Gamma Factory beam tests [28] confirmed the sufficient stability of the PSI beams in the LHC. The LHC PSI beam would have to be refilled with approximately 10 hours periodicity[17], assuring the duty cycle of the photon-source operation of more than 80%[18]. The maximal power of the GF photon beam is limited mainly by the RF power of the LHC cavities, which is needed to restore the energies of the PSI-beam particles transferred to the emitted fluorescence photons. The existing LHC RF cavities, operating at 16 MV circumferential voltage, deliver the necessary power to produce the megawatt power of photon beams.

In the following, we present three concrete scenarios for producing megawatt-class photon beams in the requisite energy range. Each of the three scenarios could re-use the laser system that will be constructed and tested in the future Gamma Factory Proof-of-Principle (GF-PoP) experiment [29] at the SPS. This system is based on a commercial laser oscillator amplified to 60 W. Laser photons are stacked in an external enhancement optical cavity with a two-mirror hemispherical geometry allowing the storage of 200 kW of average power, corresponding to 5 mJ photon pulses recirculated at a 40 MHz rate. For more details on the GF laser system see Ref. [30].

In each of the scenarios presented below, the partially stripped ions driving the Gamma Factory photon beams are chosen to be helium-like ions carrying two electrons. This choice allows us to control the polarization of the Gamma Factory photons.

#### 1. Helium-like tin scenario

The laser-photon wavelength to resonantly excite the $1s^2\,{}^1S_0 \to 1s2p\,{}^1P_1$ atomic transition of the helium-like tin beam particles, $^{120}_{50}\text{Sn}^{48+}$, is chosen to be $\lambda = 517$ nm. This photon wavelength can be obtained by doubling the frequency of the photons produced by the GF-PoP $\lambda = 1034$ nm laser. In this scenario, the relativistic Lorentz factor $\gamma_L$ of the beam particles which satisfies the resonance condition for head-on collisions of the $^{120}_{50}\text{Sn}^{48+}$ ions with the laser pulses is $\gamma_L^{\text{Sn}} = 5362$. It corresponds to the equivalent proton energy of 12.47 TeV. Such energy cannot be reached with the present LHC dipoles but can be realised in the future HE-LHC project [14]. Photon beams of power up to 2.8 MW with the maximal photon energy of 275.8 MeV can be generated by the $^{120}_{50}\text{Sn}^{48+}$ beam with the presently operating LHC cavities and 1000 bunches of $10^9$ ions per bunch circulating in the LHC ring.

#### 2. Helium-like xenon scenario

The laser-photon wavelength to resonantly excite the $1s^2\,{}^1S_0 \to 1s2p\,{}^1P_1$ atomic transition of helium-like xenon beam particles, $^{129}_{54}\text{Xe}^{52+}$, is chosen to be $\lambda = 258.5$ nm. Photons at this wavelength can be produced by two consecutive frequency doubling stages of the GF-PoP $\lambda = 1034$ nm laser photons. In this scenario, the relativistic Lorentz factor $\gamma_L$ of the beam particles which satisfies the resonance condition is $\gamma_L^{\text{Xe}} = 3149$. It corresponds to an equivalent proton energy of 7.27 TeV, which is slightly above the allowed LHC-beam momentum range, reflecting the present quench-protection limit of the maximal current of the LHC dipoles. The maximal photon-beam energy that could be reached in this scheme is 190.2 MeV. Photon beams of the power up to 3 MW can be generated by the $^{129}_{54}\text{Xe}^{52+}$ beam with the presently operating LHC cavities and 1000 bunches of $10^9$ ions per bunch. Due to the reduced $\Delta$-resonance cross section for 200 MeV photons, the higher cross section for pion absorption in the target material and, finally, excessive (relative) pion energy loss in the photon target, the number of pions produced by the photon beam is expected to be smaller by a factor of almost 10 than in the helium-like tin scenario, assuming the same photon beam power.

#### 3. Helium-like ytterbium scenario

A possible remedy to increase the photon beam energy is to use ions with a higher atomic number $Z$. This can be done at the cost of further reducing the laser photon wavelength – to the value of $\lambda = 129.25$ nm. Photons of such a wavelength can be produced by performing three consecutive frequency doubling stages of the $\lambda = 1034$ nm photons generated by the GF-PoP laser. They can resonantly excite the $1s^2\,^1S_0 \to 1s2p\,^1P_1$ atomic transition of helium-like ytterbium beam particles, $^{174}_{70}\text{Yb}^{68+}$, provided that the relativistic Lorentz factor $\gamma_L$ of the beam particles is chosen to be $\gamma_L^{\text{Yb}} = 2731$. It corresponds to the equivalent proton energy of 6.47 TeV. This value is within the allowed LHC-beam momentum range. The maximal photon beam energy to be reached in this scheme is 286.2 MeV. A GF photon beam of power up to 4 MW can be generated in this scheme with the presently operating LHC cavities and 1000 bunches of $10^9$ ions per bunch.

---

[17] The electron stripping by the residual gas in the LHC rings is the main process which limits the PSI-beam lifetime.
[18] For operation at high beam-intensities, collimation and cooling of the PSI beam with a dedicated laser system may be required.

### C. Feasibility

Important technical challenges for a concrete realisation of each of the above three scenarios must be addressed prior to the choice and implementation of the optimal one. Complementary studies and/or technical developments are necessary to asses their technical feasibility.

Beams of tin and ytterbium have never been produced at CERN. Studies of the corresponding ion sources, the electron stripping stages as well as the acceleration strategy in LEIR, PS, and SPS would need to be made, and the maximal bunch-charge limits would have to be evaluated.

Reaching the PSI bunch intensity of $10^9$, corresponding to a canonical ion-beam operation at RHIC, may be difficult for the CERN ion-injection system, in particular for the highest-$Z$ ions. The discussion of this aspect can be found in Ref. [14]. This challenge is not critical as the limits of the bunch population at the early acceleration stages of the PSI beams can be circumvented by slip-stacking of the injector's bunches at the LHC injection energy, followed by the laser cooling. A complementary remedy would be to multiply the number of interaction points of the laser photons with the PSI bunches.

The first two of the above three scenarios are only possible following an upgrade of the integrated bending power of the LHC dipoles, e.g. the helium-like tin scenario can be realised only if the HE-LHC upgrade [14] is made in the future. The third scenario requires a dedicated R&D effort to extend the "reflectivity range" of the Fabry–Pérot (FP) cavity mirrors down to the $\lambda = 130$ nm range, or technological progress in free-electron-laser (FEL) or free-ion-laser (FIL) focused on the reduction of their bandwidth and the increase of their repetition rate to the MHz range. Finally, upgrading the GF photon source power to 10 MW would require adding new RF cavities to those presently installed. For the third scenario, three new cavities of the same circumferential voltage would have to be added.

All the above aspects require dedicated studies which go beyond the scope of this initial, exploratory paper. Such studies should be pursued if the exploratory studies of the GF-driven muon source demonstrate its significant potential to be competitive or even more attractive than the present and future proton-driven sources. To quantitatively evaluate such a potential we concentrate, in the following, on the third of the above scenarios – the helium-like ytterbium one. We present detailed simulations of the corresponding GF photon beam and its capacity to produce high-intensity muon and positron beams. The photon fluxes and their characteristics for the other two scenarios, assuming the same laser-pulse power and the IP optics, are similar to the chosen case.

We start with the presentation of the dedicated Gamma Factory simulation tools that we have developed and used for the helium-like ytterbium scenario case studies.

### D. Tools and simulation parameters

Interactions of the PSI bunch with the laser pulse including processes of atomic photon absorption, spontaneous and stimulated photon emissions have been simulated using the Monte Carlo event generator **GF-CAIN** [29, 31] for a helium-like ytterbium-ion beam with the parameters given in Table IV. **GF-CAIN** is a customised version of the Monte Carlo program **CAIN** [32] developed at KEK-Tsukuba, Japan, for the ILC project, dedicated to the Gamma Factory.

Since the lifetime of the excited state $\tau' = 1.01 \times 10^{-16}$ s is much shorter than the laser-light pulse duration in the ion-rest frame $\sigma_t' \simeq 915 \times 10^{-16}$ s, the two-level atomic system has enough time to reach a steady state after an intermediate period of Rabi oscillations, and thus interactions of these helium-like Yb ions with laser light can be described within a probabilistic, scattering cross-section formalism [33]. This, in turn, can be a basis for developing an appropriate Monte Carlo algorithm and constructing an event generator that would directly simulate the corresponding process.

The process of a PSI-bunch interaction with laser-light pulse includes the atomic photon absorption by the Yb ion in its ground state to jump to the excited state, and then the subsequent spontaneous (after a finite excited-state lifetime) or stimulated photon emission to return to the ground state. Details of the corresponding Monte Carlo algorithm and event generation are given in Ref. [33]. Because the lifetime of the considered excited state of the He-like Yb ion is much shorter (by a factor of $\sim 900$) than the laser-pulse duration, the ion can be excited many times in the course of its interaction with the laser pulse, resulting in multiple photon emissions. For the beam parameters given in Table IV, each ion can emit spontaneously on average $\sim 20$ photons during every passage of the laser-light pulse through the PSI-beam bunch[19]. The maximum energy of the emitted photons,

---

[19] The quoted number is for the zero collision angle of the PSI bunches and the laser pulses. For IP optics with a nonzero crossing angle, the reduction of the number of emitted photons from

TABLE IV: Input parameters for the **GF-CAIN** simulations with the He-like Yb scenario.

| PSI beam | $^{174}_{70}\text{Yb}^{68+}$ |
|---|---|
| $m$ – ion mass | $161.088\,\text{GeV}/c^2$ |
| $E$ – mean energy | 440 TeV |
| $\gamma_L = E/mc^2$ – mean Lorentz relativistic factor | 2731.3 |
| $N$ – number ions per bunch | $10^9$ |
| $\sigma_E/E$ – RMS relative energy spread | $2 \times 10^{-4}$ |
| $\beta_x = \beta_y$ – $\beta$-function at IP | 0.5 m |
| $\sigma_x = \sigma_y$ – RMS transverse size | 16 $\mu$m |
| $\sigma_z$ – RMS bunch length | 15 cm |
| Bunch repetition rate | 20 MHz |
| Laser | Yb:YAG |
| $\lambda$ – photon wavelength | 129.25 nm |
| $\hbar\omega$ – photon energy | 9.5926 eV |
| $\sigma_\lambda/\lambda$ – RMS relative band spread | $2 \times 10^{-4}$ |
| $U$ – single pulse energy at IP | 5 mJ |
| $\sigma_x = \sigma_y$ – RMS transverse intensity distribution at IP | 20 $\mu$m |
| $\sigma_z$ – RMS pulse length | 15 cm |
| $\theta_l$ – collision angle | 0 deg |
| Atomic transition of $^{174}_{70}\text{Yb}^{68+}$ | $1s^2\,^1S_0 \to 1s2p\,^1P_1$ |
| $\hbar\omega'_r$ – resonance energy | 52.4 keV |
| $\tau'$ – mean lifetime of spontaneous emission | $1.01 \times 10^{-16}$ s |
| $g_1, g_2$ – degeneracy factors of the ground and excited states | 1, 3 |
| $\hbar\omega_1^{\max}$ – maximum emitted photon energy | 286.2 MeV |

in this case, is $\sim 286.2\,\text{MeV}$.

In our studies, we have included, for the first time, the photon polarization degrees of freedom. They are implemented in **GF-CAIN** according to Ref. [36]. Photon polarizations are described by three Stokes parameters: $P_1, P_2$ and $P_3$, where according to the convention of Ref. [36], $P_1$ and $P_2$ correspond to linear polarizations, while $P_3$ corresponds to the circular polarization. In the studies presented in this paper, we consider only the circular polarization which in terms of the Stokes parameter $P_3$ is described as follows:

$$P_3 = \frac{I_+ - I_-}{I_+ + I_-}, \qquad (14)$$

where $I_+$ and $I_-$ are the intensities of the right-handed and left-handed circularly polarized light, respectively. $P_3 = +1(-1)$ corresponds to a fully right(left)-handed

---

the PSI bunch can be compensated by cloning the laser systems, installing the requisite number of them over 20-meter-long straight sections of the ion beam-line and by running a large $\beta^* = 50\,\text{m}$ IP optics. This solutions has already been evaluated in the ongoing design studies of the GF-beam-driven 300 MW subcritical reactor for CERN [34], based on the concept presented in [35].

circular polarization. The studies presented in this paper are restricted to the case of fully right-handed circularly polarized laser photons, i.e. $P_1^i = P_2^i = 0$ and $P_3^i = 1$.

The angular distribution of the spontaneously emitted photons in the ion-rest frame, with $+z$ axis along the PSI-beam direction in the laboratory frame, is given by [36]

$$\frac{d\sigma}{d\Omega'_\gamma}(\theta'_\gamma, \phi'_\gamma) = \frac{\sigma}{4\pi}\,W(\theta'_\gamma, \phi'_\gamma), \qquad (15)$$

where

$$W(\theta'_\gamma, \phi'_\gamma) = \frac{3}{4}\left(1 + \cos^2\theta'_\gamma\right), \qquad (16)$$

and the Stokes parameter $P_3^f$ for the final-state (emitted) photon is given by

$$P_3^f\left(\theta'_\gamma, \phi'_\gamma\right) = \frac{-3P_3^i \cos\theta'_\gamma}{2\,W(\theta'_\gamma, \phi'_\gamma)}. \qquad (17)$$

The photon polarization degrees of freedom does not play an important role for the muon source based on the decays of the resonantly produced pions. They become, however, important for leptons produced by the conversion of photons in the electromagnetic field of the target atoms.

In **GF-CAIN**, the angular distribution of the emitted photon is generated according to the distribution $W\left(\theta'_\gamma, \phi'_\gamma\right)$ of Eq. (16) and the Stokes parameter $P_3^f$ is computed according to Eq. (17) in the ion-rest frame, and then the photon four-momentum is Lorentz-transformed to the laboratory (LAB) frame. The Stokes parameter $P_3^f$ is Lorentz-transformation-invariant, see e.g. [37], and, in particular, it does not change when the photon emission angles are expressed in the LAB frame rather than in the ion-rest frame. Angular distributions of the Stokes parameters in the laboratory frame generated in **GF-CAIN** were cross-checked against analytical formulae given in Ref. [36] and a very good agreement was found [38].

In our initial simulations, we have assumed that the ion-bunch parameters given in Table IV are restored within the bunch revolution time in the LHC and are unchanged for each consecutive crossing with the laser pulse. As already discussed, to fulfil this condition, the RF power of the LHC cavities should be large enough to restore the energy lost in the process of photon emission. If necessary, stabilisation of the emittance of the PSI bunches can be achieved with a dedicated reduced-bandwidth laser system, such as the one described in Ref. [39].

### E. Simulation results

In Fig. 21 we present the correlations of the emitted photon energy $E_\gamma$ in the LAB frame with (1) its emission angle $\theta_\gamma$ w.r.t. the PSI beam direction Fig. 21a and (2) with its circular-polarization Stokes parameter $P_3^f$ Fig. 21b for the fully right-handed circularly polarized laser photons ($P_3^i = 1$). Fig. 21a shows the canonical feature of the GF photon source: the energy of the emitted photons is directly correlated with its emission angle. The width of this correlation depends upon the emittance of the PSI beam and on the optics of the ion–photon interaction point (IP). In the presented studies, the canonical parameters of the typical LHC beams and of the IP design were chosen: $\beta^* = 0.5\,\mathrm{m}$ and the normalised beam emittance $\epsilon_N = 1.4\,\mathrm{mm} \times \mathrm{mrad}$. Fig. 21b shows another canonical feature of the GF photon source: the polarization of the emitted photons is directly correlated with the photon energy (emission angle). The initial laser photon polarization is preserved for photons emitted in the direction of the incoming laser photons and reversed for photons emitted in the direction of the incoming ions.

Projections of the energy–angle correlation plot on the $E_\gamma$ and $\theta_\gamma$ axes are shown in the (a) and (b) panels of Fig. 22, respectively. Our choice of closed-shell helium-like ions and circularly polarized laser photons maximises the energy peak close to the maximum energy of the photons. The angular divergence of the GF photon source is driven by the $\gamma_L$-factor of the PSI beam. Photons in the high-energy region are emitted, for LHC-beam energies, in a very narrow cone and give rise to a well-collimated photon beam (half of the most energetic photons are emitted with the angle $\theta_\gamma \leq 1/\gamma_L$).

In Fig. 23a, the $x$–$y$ coordinates distribution of the emitted photons is shown at a distance of $50\,\mathrm{m}$ from the photon production point. This plot is centred at the crossing of the line tangential to the incoming ion-beam direction and the plane perpendicular to this direction. If the photon beam target is placed at $50\,(100)\,\mathrm{m}$ distance, about 50% of the photon beam power would be confined within a circle of $1\,(2)\,\mathrm{cm}$ radius. Fig. 23b shows that within a beam-spot of $1\,\mathrm{cm}$ radius at $z = 50\,\mathrm{m}$, the photons are circularly polarized. The average photon polarization, as specified by the Stokes parameter $P_3$, reaches a value of $P_3 = -0.95$.

As discussed in Subsection V B, the energy bandwidth of the photon beam should not exceed the bandwidth of the Fermi motion of nucleons within nuclei, if one wants to preserve the high spectral density of the produced pions – as in the case of monochromatic photon beams. The width of the photon energy for the GF photon source is determined, at the fixed transverse PSI beam-spot size, by the $\beta^*$ value of the PSI beam and by the size of the radial collimation slit of the photon beam. In Fig. 24, we present the photon energy distributions for two values of the $\beta^*$ parameter and two scenarios of the radial collimation of the beam. Changing the $\beta^*$ value from $0.5\,\mathrm{m}$ to $5\,\mathrm{m}$ could reduce the width of the energy distribution. If the PSI-beam transverse spot is preserved, such a change would require reducing of beam transverse emittance by efficient laser beam cooling, as discussed in detail in Ref. [39].

The integrated beam power for photons contained in a beam-spot of radius 5 and $10\,\mathrm{mm}$ at a $50\,\mathrm{m}$ distance from the IP is given in each of the corresponding plots of Fig. 24. It ranges from $1.64\,\mathrm{MW}$ to $4.47\,\mathrm{MW}$, depending of the chosen photon beam collimation slit and the $\beta^*$ value. The plots in Fig. 24 illustrate the tuning capacity of the optimal GF photon-beam parameters to maximise the number of pions produced and their spectral density. A megawatt-class photon beam in the optimal energy range for the resonant pion production can be generated at CERN, if one of the studied scenarios is realised.

We can now finalise our exploratory work by comple-menting the studies based on a hypothetical monochromatic photon beam, presented in Section IV, with the ones based on the simulated GF photon beam, described in this section. The results of these studies are reported in the following section.

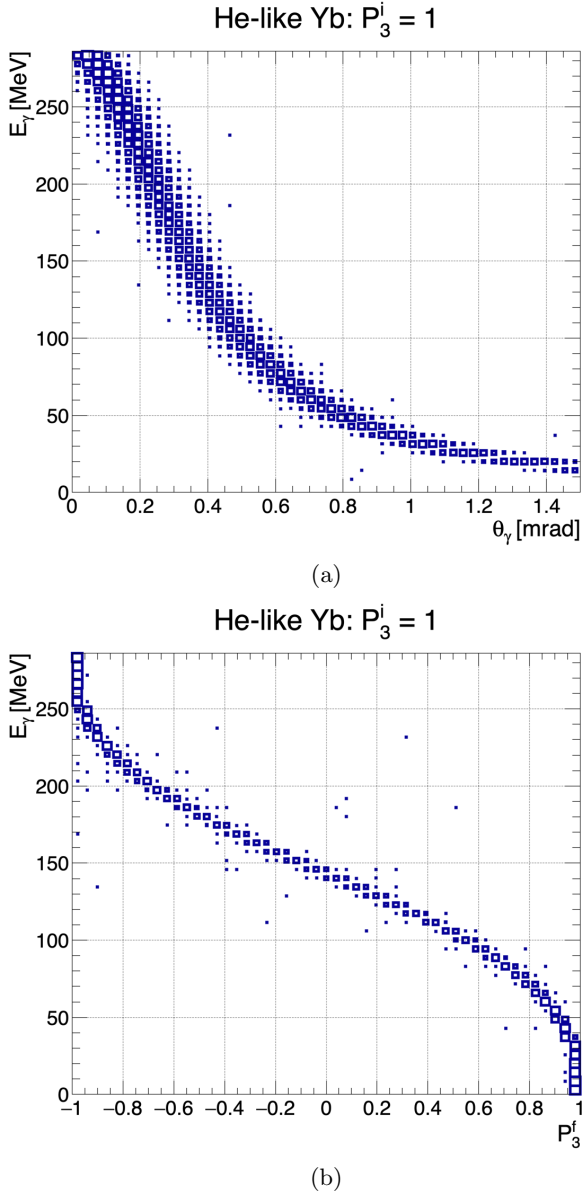

FIG. 21: Correlations of the emitted photon energy $E_\gamma$ in the LAB frame with its emission angle $\theta_\gamma$ w.r.t. the PSI beam direction (a) and with its circular-polarization Stokes parameter $P_3^f$ (b) for fully right-handed circularly polarized laser light.

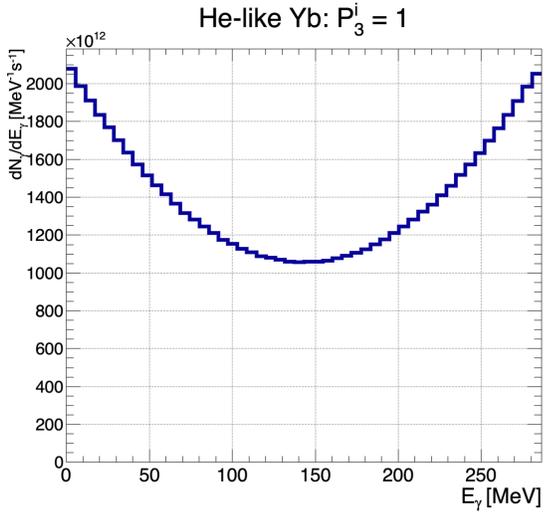

(a)

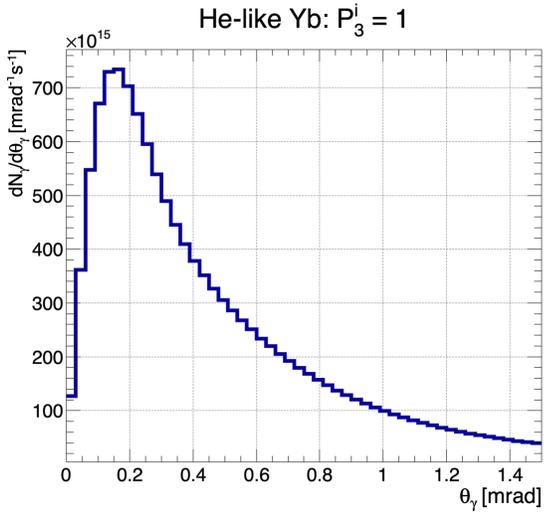

(b)

FIG. 22: The energy spectrum of the emitted photons (a) and their angular distribution (b).

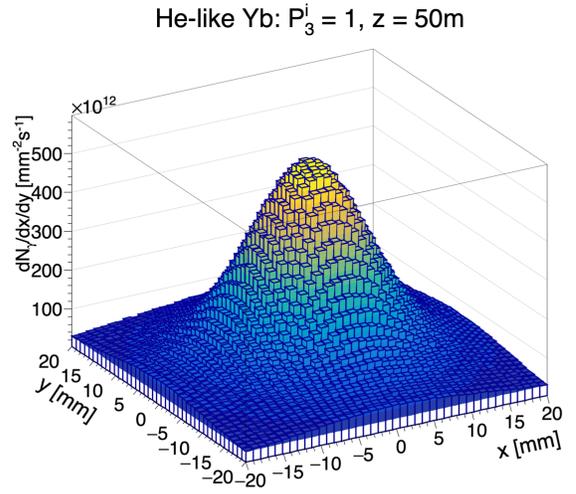

(a)

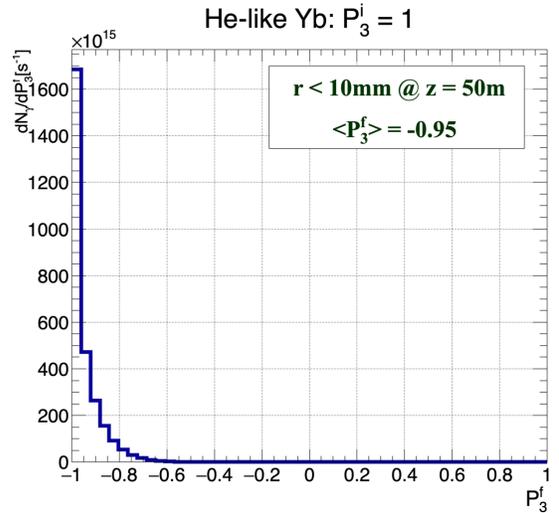

(b)

FIG. 23: The distribution of the impact points of the emitted photons on a target placed at the distance of 50 m from the interaction point (a) and the effective photon polarization corresponding to the beam with the transverse radius $r < 10$ mm (b) for the fully right-handed circularly polarized laser light.

## VI. GAMMA FACTORY LEPTON-SOURCE CHARACTERISTICS

### A. Pions

In Section IV, we have demonstrated that more than $10^{13}$ pions per second of both signs can be produced and selected in collisions of a 1 MW power 300 MeV monochromatic photon beam with a graphite target of 20 cm length. Pions produced in the processes of excitations of the $\Delta$ resonances are quasi-monoenergetic with the width of their momentum (transverse-momentum) distribution of 24 (22) MeV/c, essentially driven by the Fermi-motion smearing of the nucleon momenta in the target nuclei.

The simulated Gamma Factory photon beam is no longer monochromatic and the spectral density of pions, at a fixed beam power, is expected to deteriorate. Two main factors determine the width of the momentum (transverse-momentum) distribution of the produced pions: the size of the collimation slit for the photon beam and, to a lesser extent, the $\beta^*$-parameter value of the PSI beam at the photon-beam production point.

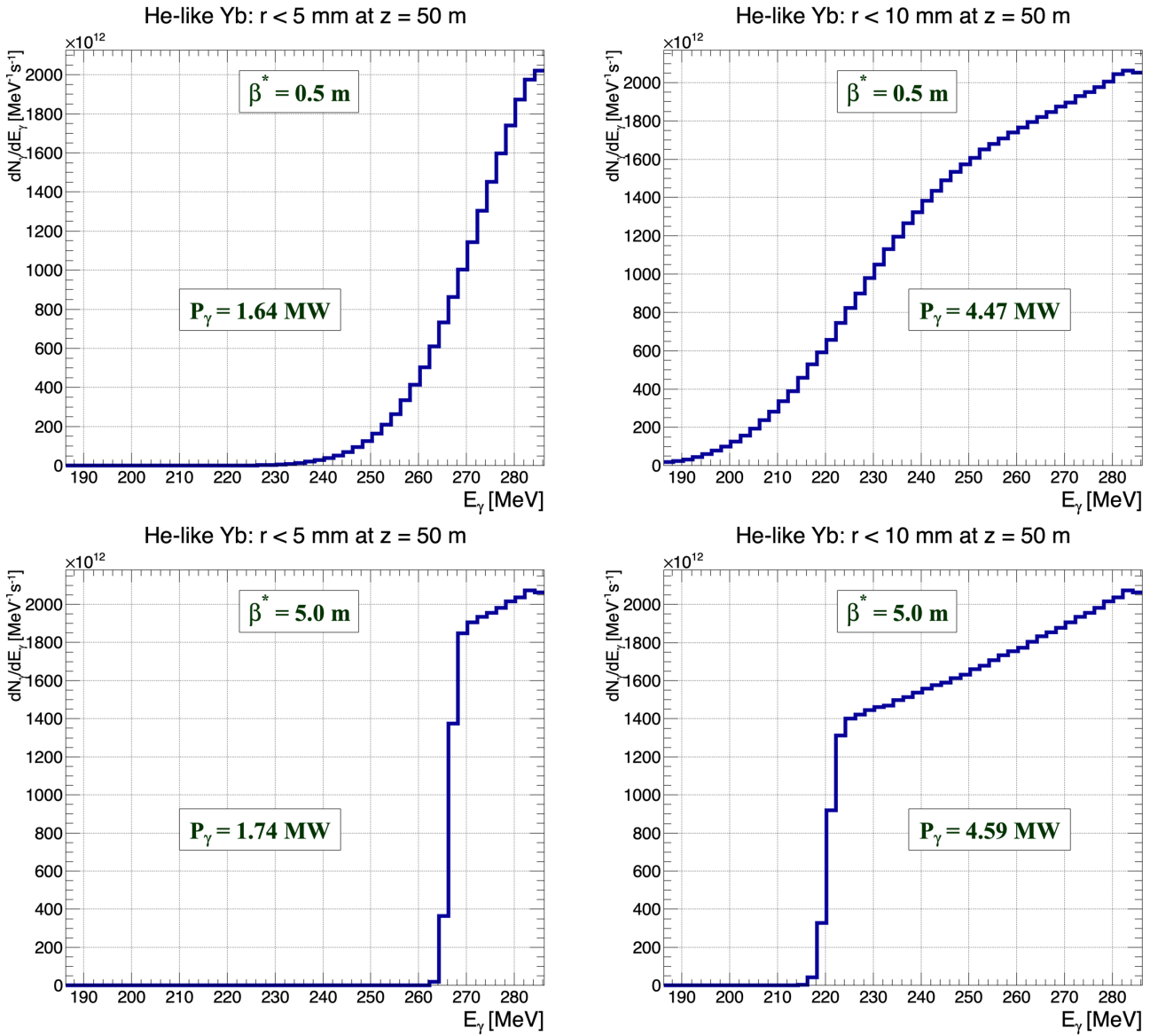

FIG. 24: The photon beam energy spectrum for various IP optics.

In Figs. 25 and 26, we show the momentum and the transverse-momentum distributions of pions coming out of a 20 cm long graphite target and fulfilling the pion selection criteria specified in Section IV. The presented plots are made for the $\beta^* = 0.5$ m scenario. Increasing the $\beta^*$ value by a factor of 10 with a simultaneous decrease of the PSI-beam transverse emittance by the same factor, allowing preservation of the PSI-beam spot size at the collision point with laser pulses, does not bring a significant improvement in the pion yields and their spectral density. The plots are made for a photon beam collimated at the distance of 50 m from its production point by a radial collimator slit of $r_\gamma \leq 5$ mm (Fig. 25a) and $r_\gamma \leq 10$ mm (Fig. 25b). Opening the radial collimator slit to the latter value increases the number of selected negative pions from $2.75 \times 10^{13}$ s$^{-1}$ to $7.13 \times 10^{13}$ s$^{-1}$ and the number of selected positive pions from $2.95 \times 10^{13}$ s$^{-1}$ to $7.79 \times 10^{13}$ s$^{-1}$. Such an increase of the pion yields is associated with a deterioration of the RMS of the momentum distribution from 31 MeV/c to 37 MeV/c and the RMS of the transverse momentum distribution from 25 MeV/c to 31 MeV/c.[20] These values are by a factor of 1.5 larger than those for the monochromatic point-like beam discussed in Section IV. For the GF photon beam, the RMS values of these distributions are no longer driven by only the Fermi motion of the nucleons in the nuclear target, but also by the accepted energy range of the GF photons. The chosen GF photon-beam production and collimation scheme maximises the spectral density of produced pions.

---

[20] In the studies presented in this section, we have assumed that the pion-production graphite target is placed just behind the photon-beam collimation slit. If the target is placed at a distance of 100 (200) m from the IP position, then the GF photon-beam spot size is by a factor of 2 (4) larger. In such a case, the radius of the target would have to be increased from 2.5 cm to 5 cm. Such a change would not cause a significant further deterioration of the pion spectral density in comparison to the monochromatic photon beam.

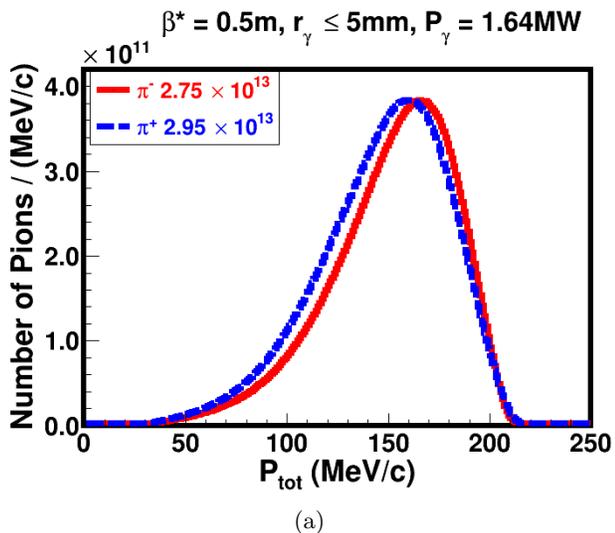
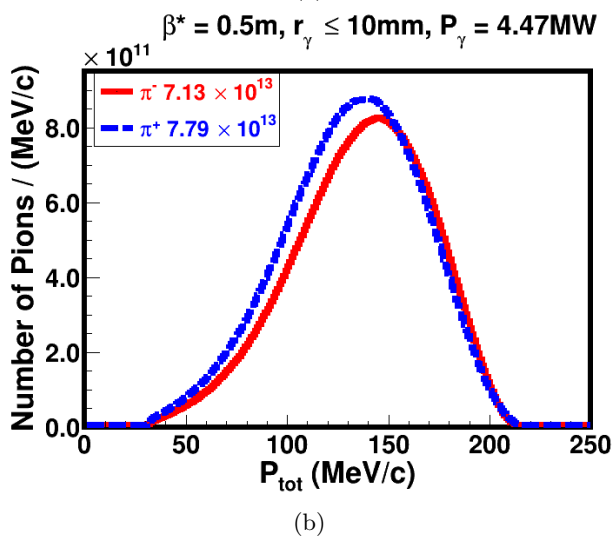

FIG. 25: Momentum distributions of pions produced by the Gamma Factory photon beam coming from collisions of a helium-like ytterbium beam with the laser pulses and collimated at a distance of 50 m from the photon-production point by a $r_\gamma \leq 5$ mm slit (a) and $r_\gamma \leq 10$ mm slit (b).

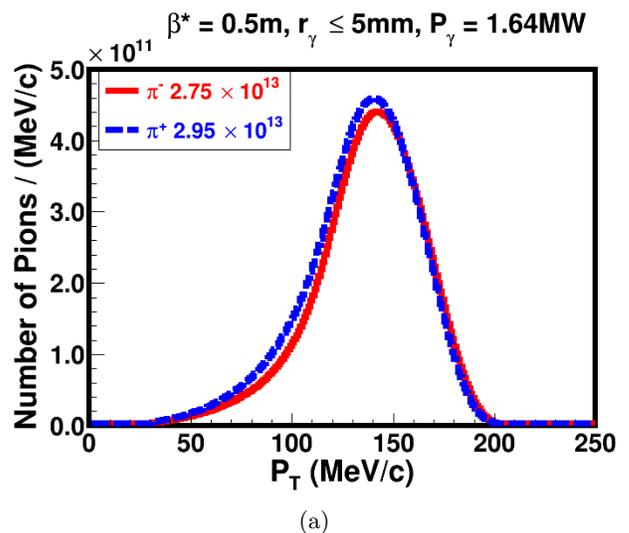
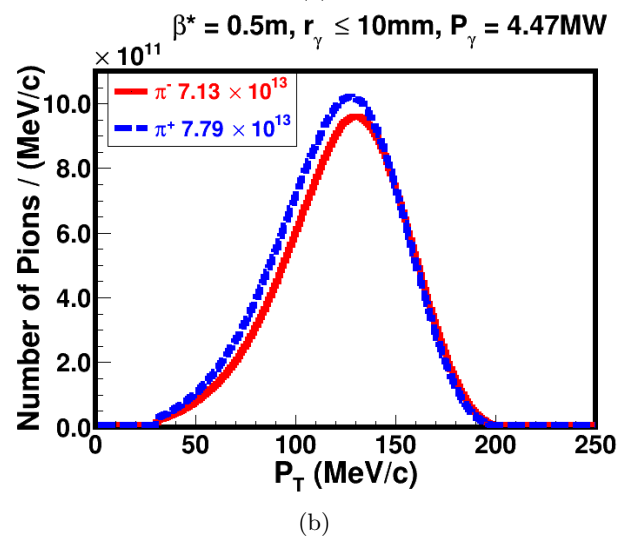

FIG. 26: Transverse momentum distribution of pions produced by the Gamma Factory photon beam coming from collisions of a helium-like ytterbium beam with the laser pulses and collimated at a distance of 50 m from the photon-production point by a $r_\gamma \leq 5$ mm slit (a) and $r_\gamma \leq 10$ mm slit (b).

The integrated beam power for photons contained in a beam spot of radius of 5 mm and 10 mm at a 50 m distance from IP is given in each of the corresponding plots. Note that while for the 5 mm collimated beam, the estimated energy deposited in the target is 237 kJ/s, the corresponding value for the 10 mm collimated one is 635 kJ/s. In the latter case, it is larger than the one that can be evacuated from the 10 cm radius target by a water cooling system (see the discussion in Section IV D). A special target would have to be designed, e.g. a horizontally rotating wheel target following the ESS-target design [40], to handle such a large energy deposition. This aspect deserves specialised studies which are beyond the scope of the present paper.

### B. Muons

Pion production is an intermediate step of the production of the muon beam. The Gamma Factory photon beam produces pions that are non-relativistic and a large majority of them decay into muons and neutrinos over their propagation path of the 20 m length, as illustrated in Fig. 27.

Fig. 28 shows the momentum distribution of muons coming from pions decaying before reaching their 20 m long propagation path. These plots are made assuming that all the pions satisfying the pion selection criteria specified in Section IV are collected. Fig. 28a corresponds to the photon beam collimated at the distance of 50 m from its production point by a radial slit of $r_\gamma \leq 5$ mm and Fig. 28b to the one collimated by a radial collimation slit of $r_\gamma \leq 10$ mm. The numbers of $\mu^-$ produced for the above two beam-collimation scenarios are $2.32 \times 10^{13}\,\text{s}^{-1}$ and $6.18 \times 10^{13}\,\text{s}^{-1}$, respectively, and the ones for the produced $\mu^+$ are $2.50 \times 10^{13}\,\text{s}^{-1}$ to $6.79 \times 10^{13}\,\text{s}^{-1}$, respectively.

These momentum-distribution plots illustrate the principal characteristic of the GF muon source: muons are produced in a narrow momentum range with a rather unprecedented transverse-momentum-integrated spectral density, reaching the value of $7 \times 10^{11}$ muons/s/MeV. The small asymmetry in the number of the positive and negative muons reflects the charge asymmetry of the produced pions, as predicted by the present GEANT4 model. Its origin remains to be understood and the size of the effect to be confirmed in a dedicated experiment.

### C. Positrons

The majority of particles that can be produced in collisions of the GF photon beam with a carbon target are electrons and positrons. Their phase space and their timing characteristics are different from those of the pions, as discussed in Section IV E. As a consequence, their col-lection scheme may be, to a large extent, independent of the pion collection scheme. The GF-beam-driven source could thus play a double role: that of a muon and of positron source.

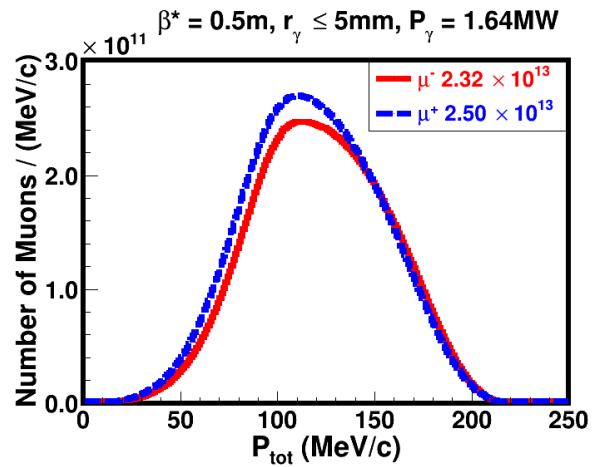

(a)

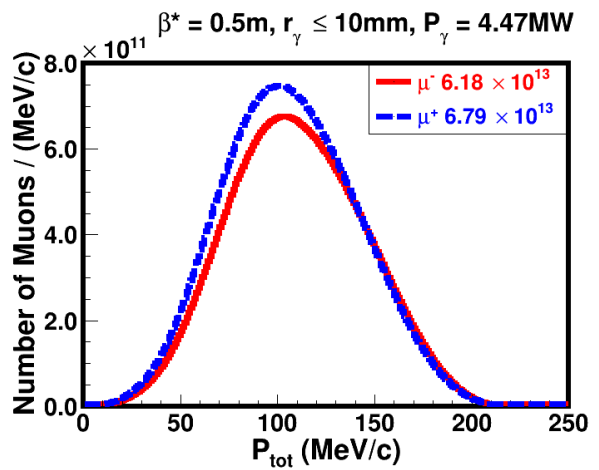

(b)

FIG. 28: Momentum distributions of muons coming from pions decaying before reaching a distance of 20 m of their propagation path. Photons are produced by collisions of a helium-like ytterbium beam with the laser pulses. They are collimated at a distance of 50 m from the photon production point by a $r_\gamma \leq 5$ mm slit (a) and a $r_\gamma \leq 10$ mm slit (b).

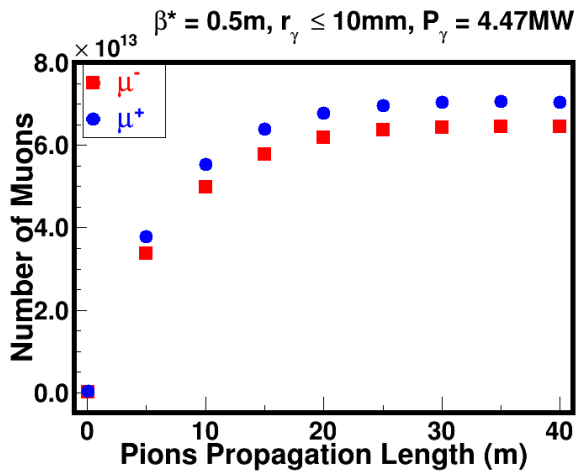

FIG. 27: The muon flux as a function of the pion propagation length. Pions are collected in the angular range of $40° - 120°$. Their transverse momentum satisfies the condition: $P_T > 30$ MeV/c. Photons are produced by collisions of a helium-like ytterbium beam with the laser pulses. They are collimated at a distance of 50 m from the photon production point by a $r_\gamma \leq 10$ mm collimation slit.

In Fig. 29, the momentum distributions of electrons and positrons at the exit of the graphite target are shown. As before, Fig. 29a corresponds to a photon beam collimated at the distance of 50 m from its production point by a radial slit of $r_\gamma \leq 5$ mm and Fig. 29b to the one collimated by the radial collimation slit of $r_\gamma \leq 10$ mm. The small asymmetry in the spectra of electrons and positrons reflects the effects of the Compton scattering of photons on the atomic shell electrons and, to a lesser extent, the disappearance of positrons due to their annihilation with the target electrons. If these processes are switched off in the GEANT4 simulation, the positron–electron symmetry is fully restored.

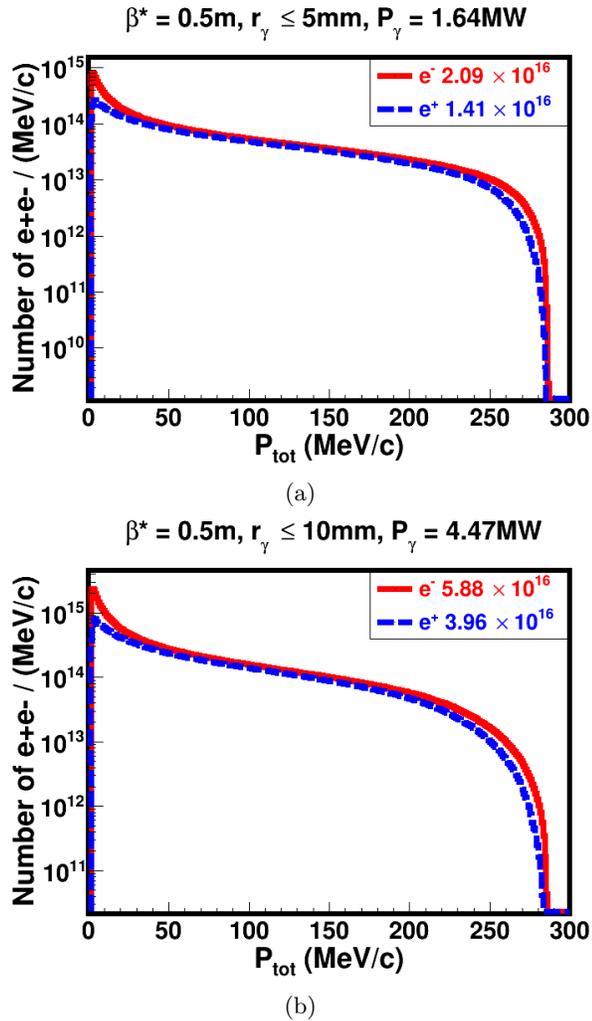

FIG. 29: Momentum distribution of electrons and positrons exiting the photon beam target. Photons are produced by collisions of a helium-like ytterbium beam with laser pulses. They are collimated at a distance of 50 m from the photon production point by a $r_\gamma \leq 5$ mm slit (a) and a $r_\gamma \leq 10$ mm slit (b).

These plots demonstrate that such a "parasitic" positron source could produce up to $3.96 \times 10^{16}$ positrons per second. The majority of high-momentum positrons come from conversions of the primary circularly polarized photons which produce longitudinally polarized positrons[21].

The scenario of the PSI beam and the laser system which was simulated and presented in this paper is optimal for a muon source. It is, obviously, not the optimal one for the GF polarized positron source. For the latter, a helium-like calcium beam and a conventional laser can be used to produce a GF photon beam with a maximal energy of $\approx 20$ MeV. As the GF photon flux is limited by the LHC RF power, the intensity of the positron source in such a scenario is expected to be improved by at least a factor of 10 at the equivalent LHC RF power, while keeping approximately the same energy deposition in the target material.

This result and the perspective of increasing the positron flux with a lower-energy photon beam is encouraging – the GF fluxes could potentially satisfy the requirements of future high-energy electron–positron colliders and the LEMMA-scheme muon colliders [8].

### D. Timing structure

Gamma Factory photons can be produced in sub-nanosecond-long pulses with a 20 MHz repetition rate. Such a timing structure would reflect the present bunch-filling pattern and the bunch length of ions stored in the LHC ring. A muon source, driven by such a beam, would be essentially a continuous (CW) source. While such a source fulfils the requirements of experiments looking for rare muon decays and/or the requirements of the muon-beam-based neutrino factory, its efficient use for a muon collider is not straightforward. The optimal muon source for the muon collider should deliver a comparable number of muons per second as the GF-beam-driven one, but in a small number of bunches[22].

In order to adapt the Gamma Factory muon source to the needs of the muon collider, an efficient bunch-merging scheme – initially, for the PSI bunches at the LHC, and subsequently, for the muon bunches at the muon collider top energy – which preserves the low emittance of the muon beam would have to be developed. Another possibility would be to change the collider–detector design paradigm by developing a long detector capable of observing collisions of a group of bunches with a group of counter-propagating bunches. The discussion of all these options requires dedicated studies and is beyond the scope of the work presented in this paper.

## VII. NEXT STEPS

The studies presented in the previous section can be considered as an initial, exploratory step. Conceptual and technical designs of the pion, muon, and positrons collection scheme would have to be developed to find out which fraction of produced and selected particles would finally be transformed into bunches of muon and positron beams. Such studies have been performed over the last 30 years for proton-beam-driven muon sources, see e.g. Ref. [41]. They need to be initiated for the photon-beam-driven sources, e.g. by considering the GF scheme proposed in this paper.

---

[21] The momentum distributions of electrons are shown in Fig. 29 to indicate the contribution of the processes of Compton scattering of the GF photons on the carbon atomic shells electrons.

[22] For example, for a 10 TeV muon collider, the preferred scheme is to collect the equivalent number of mouns in bunches produced with 10 Hz frequency.

The efficient collection of pions, and subsequently muons, over a distance of $\approx 20$ m is highly non-trivial. The initial idea to be studied is to profit from the small RMS values of the longitudinal and transverse-momentum distributions and design the pion collection zone with the requisite toroidal magnetic field and an accelerating-gradient electric field to collect the majority of the positive (negative) pions produced over a large angular-acceptance region. The toroidal magnetic field in the pion-production zone should focus pions momenta such that, on average, their momentum vectors are parallel to the initial photon propagation axis. This rotation should be done prior to forming the pion beam with the solenoidal magnetic field.

The pion-beam forming stage has to be quick enough to collect the majority of muons produced by the pion beam already moving parallel to the photon beam axis. Only in such a case, can the polarization of the muons be controlled. In the toroid scheme of pion collection, two independent GF photon beams would have to be created to separately collect the positive and negative pions and muons.

## VIII. CONCLUSIONS

In this paper, we have presented exploratory studies of the photon-beam-driven muon and positron source. We have proposed three Gamma Factory scenarios to produce megawatt-class photon beams which are optimal for muon production. Then, we have performed detailed simulations of one of these scenarios: the one based on helium-like ytterbium beams stored in the LHC.

We have optimised the beam and target parameters, and compared the photon-beam-driven scheme with the canonical proton-beam-driven one. We have demonstrated that the GF scheme has a potential of increasing the intensity of the presently operating muon and positron sources by up to 4 orders of magnitude, generating fluxes larger than $10^{13}$ muons and $10^{16}$ positrons per second.

The GF muon source proposed in this paper – based on the exclusive single-pion production process – maximises the spectral density of the produced pions. This feature may facilitate the design of the pion/muon collection scheme to produce low-emittance muon beams and could allow control of the degree of longitudinal polarization of the collected muons. Polarized positrons, produced parasitically, can fulfil the requirements of the LEMMA positron source.

The exploratory studies presented in this paper constitute only an initial small step in the process of developing a concrete technical design of the photon-beam-driven source. Their goal is to encourage further "expert" studies. There are three principal reasons why in our view such studies should be pursued: (i) the photon-beam-driven source has several potential merits that proton-beam-driven sources do not have, (ii) while the proton accelerators – which are capable of delivering the requisite megawatt-power proton beams – remain to be designed and/or constructed, a comparable megawatt-power photon beam can be generated at the existing storage ring, LHC, with unprecedentedly high plug-power efficiency by implementing commercially available laser systems, (iii) if the scientific life of the LHC is extended beyond its hadron-collider function – by including its GF operation as a unique quality photon source for many branches of science – then the GF lepton-source scheme could appear to be the most efficient, low-cost solution to provide the requisite, high-intensity polarized lepton beams for future accelerators and experiments at CERN.

## ACKNOWLEDGMENTS

We would like to acknowledge the support of the CERN Physics Beyond Colliders framework providing a cradle for the ongoing Gamma Factory studies. Special thanks to Gianluigi Arduini, Brennan Goddard and Mike Lamont for their continuous support, and to the CERN GEANT team for numerous discussions and their help.

This project has received partial funding from the European Union Horizon 2020 Research and Innovation Programme under the Grant Agreement No. 101004730 (iFAST).

The research of WP has been supported in part by a grant from the Priority Research Area (DigiWorld) under the Strategic Programme Excellence Initiative at Jagiellonian University.